\documentclass[prx,twocolumn,showpacs,superscriptaddress,10pt]{revtex4-1} 
\usepackage{amsmath,amssymb,graphicx}

\usepackage{color,times}

\definecolor{darkblue}{rgb}{0, 0, 0.8}
\usepackage[colorlinks=true, breaklinks=true, linkcolor=darkblue, citecolor=darkblue, urlcolor=darkblue]{hyperref}
\newcommand{\ket}[1]{\left| #1\right\rangle}
\newcommand{\bra}[1]{\left\langle #1\right|}

\newcommand{\rs}{\rm \scriptscriptstyle}

\newcommand{\acom}[2]{\{#1,#2\}}
\newcommand{\sign}{\operatorname{sign}}
\newcommand{\com}[2]{\left[#1,#2\right]}

\begin{document}

\title{Realization of a density-dependent Peierls phase
in a synthetic, spin-orbit coupled Rydberg system}

\author{Vincent Lienhard}
\email[These authors contributed equally to this work.]{}
\author{Pascal Scholl}
\email[These authors contributed equally to this work.]{}
\affiliation{Universit\'e Paris-Saclay, Institut d'Optique Graduate School, CNRS, Laboratoire Charles Fabry, 
91127 Palaiseau Cedex, France}
\author{Sebastian Weber}
\affiliation{Institute for Theoretical Physics III and Center for Integrated 
Quantum Science and Technology, University of Stuttgart, 70550 Stuttgart, Germany}
\author{ Daniel Barredo}
 \author{Sylvain de L\'es\'eleuc}
\affiliation{Universit\'e Paris-Saclay, Institut d'Optique Graduate School, CNRS, Laboratoire Charles Fabry, 
91127 Palaiseau Cedex, France}
\author{Rukmani~Bai}
\author{Nicolai~Lang}
\affiliation{Institute for Theoretical Physics III and Center for Integrated 
Quantum Science and Technology, University of Stuttgart, 70550 Stuttgart, Germany}
\author{Michael Fleischhauer}
\affiliation{Department of Physics and Research Center OPTIMAS,
University of Kaiserslautern, 67663 Kaiserslautern, Germany}
\author{Hans Peter B\"uchler}
\affiliation{Institute for Theoretical Physics III and Center for Integrated 
Quantum Science and Technology, University of Stuttgart, 70550 Stuttgart, Germany}
\author{Thierry Lahaye}
\author{Antoine Browaeys}
\affiliation{Universit\'e Paris-Saclay, Institut d'Optique Graduate School, CNRS, Laboratoire Charles Fabry, 
91127 Palaiseau Cedex, France}

\begin{abstract}
We  experimentally realize a Peierls phase in the hopping amplitude 
of excitations carried by Rydberg atoms, and observe the resulting characteristic chiral 
motion in a minimal setup of three sites. 
Our demonstration relies on the intrinsic spin-orbit coupling of the dipolar exchange 
interaction combined with time-reversal symmetry 
breaking by a homogeneous external magnetic field.  
Remarkably,  the phase of the hopping amplitude between 
two sites strongly depends on the occupancy of the third site, 
thus leading to a correlated hopping associated to a density-dependent Peierls phase. 
We experimentally observe this density-dependent hopping and 
show that the excitations behave as anyonic particles 
with a non-trivial phase under exchange.  
Finally, we confirm the dependence of the Peierls phase 
on the geometrical arrangement of the Rydberg atoms.
\end{abstract}

\maketitle 

\section{Introduction}

Synthetic quantum systems, i.e. well-controlled systems 
of interacting particles, are appealing to study 
many-body phenomena inspired by condensed matter 
physics~\cite{Georgescu2014}. 
One of the current challenges using this approach 
is to investigate the interplay between the non-trivial topology 
of a band-structure, resulting from, e.g., an effective magnetic field, 
and the interactions between the particles~\cite{Bergholtz2013,Cooper2019}. 
An effective magnetic field can be simulated by implementing
complex hopping amplitudes $t e^{i\varphi}$ 
between the sites of an array, characterized by a Peierls phase 
$\varphi$~\cite{Hofstadter1976,Jaksch_2003,Goldman_2014}. 
A particle circulating around a closed loop then acquires a 
phase analog to the Aharonov-Bohm phase, which is proportional to the 
enclosed magnetic flux. Effective magnetic fields
and complex-valued hopping amplitudes have 
been implemented on ultra-cold atom-based 
platforms~\cite{Dalibard_2010,Spielman_2013,Zhai_2015,Aidelsburger2018,Cooper2019}, 
by using laser-assisted tunneling  in an optical 
superlattice~\cite{Aidelsburger_2011}, high-frequency driving of 
a lattice~\cite{Kolovsky2011,Struck_2012,Jotzu2014}, 
and implementing synthetic dimensions~\cite{Mancini_2015,Stuhl_2015,Nascimbene2020}. 
Alternative platforms have also emerged such 
as superconducting qubits where complex-valued hopping amplitudes 
were demonstrated~\cite{Roushan_2016}, and photonic~\cite{ReviewPhotonics}  
or phononic~\cite{ReviewPhononics} systems operating  so-far in the non-interacting regime.
Here, we present the experimental realization of  
Peierls phases using the intrinsic spin-orbit coupling present in 
dipolar exchange interactions between Rydberg atoms.

Platforms involving individual Rydberg atoms are 
promising candidates to realize strongly interacting synthetic quantum 
matter~\cite{Saffman_2010,Weimer_2010}. 
The assembly of up to around 100 atoms 
in tunable geometries has already been 
achieved~\cite{Miroshnychenko_2006,Barredo_2016,Endres_2016,Kim_2016,Barredo_2018,Mello2019}. 
The two different regimes of interaction,  van der Waals and resonant 
dipole-dipole~\cite{Browaeys_2016}, have been used respectively 
to implement Ising-like~\cite{Labuhn_2016,Bernien_2017,Kim_2018} 
or XY spin Hamiltonians~\cite{Barredo_2015, deLeseleuc_2019}. 
In the resonant dipole-dipole regime, when the Rydberg atoms can be considered as two-level 
systems with states $nS$ and $nP$, the interaction results in the hopping of the $nP$ 
excitation between two sites, 
making it possible to explore transport phenomena. 
We recently used this fact to realize a symmetry protected topological 
phase for interacting bosons~\cite{deLeseleuc_2019}. 
Going beyond this two-level configuration, it has been proposed to engineer situations 
where the effective particle features an internal degree of freedom. 
There, the dipole-dipole interaction couples
this internal degree of freedom with the motional one,  
resulting in an intrinsic spin-orbit coupling \cite{Syzranov2014}. 
In combination with breaking of the time reversal symmetry, this can lead to
topological band structures characterized 
by non-zero Chern numbers~\cite{Peter_2015,Kiffner_2017,Weber_2018}. 

In this paper, we demonstrate this intrinsic spin-orbit coupling in a 
minimal setup of three Rydberg atoms in a triangle. 
A combination of static magnetic and electric fields perpendicular to the triangle allows us to isolate  
two levels in the $nP$ manifold, thus giving rise to an excitation
with two internal states. The external magnetic field naturally breaks the time-reversal symmetry, 
which, combined with 
the spin-orbit coupling, leads to a  characteristic chiral motion for a single excitation. 
We experimentally demonstrate this chiral motion and 
show that the dynamics is reversed by inverting the direction of the magnetic field.  
The chiral motion is well understood in an effective description, 
where one internal state of the excitation is adiabatically eliminated. 
In this case, the effective Hamiltonian is described by a non-trivial Peierls phase $\varphi$  
in the hopping amplitude, corresponding to a finite magnetic flux through the triangle. 
Remarkably, in this approach the Peierls phase depends 
on the absence or presence of a second excitation, and naturally gives rise to density-dependent hoppings, 
which are required for the creation of dynamical gauge fields \cite{Wiese2013},
as recently realized for ultracold atoms in optical lattices~\cite{Esslinger_2019,Clark2018}. 
Here, we demonstrate this density-dependent hopping by observing 
the absence of chiral dynamics for two excitations. 
Following \cite{Fradkin91}, the density-dependent hopping can be mapped to a hard-core anyon model with a statistical exchange angle $3 \varphi$. 
Finally, we demonstrate the ability to tune the effective magnetic flux through 
the triangle by varying the geometrical arrangement of the three atoms.
We conclude by discussing the implications of this spin-orbit coupling
on square and honeycomb plaquettes.  

\section{Spin-orbit coupling using dipolar exchange interactions}\label{sec:spinorbit}

\begin{figure}[t!]
\centering
\includegraphics[]{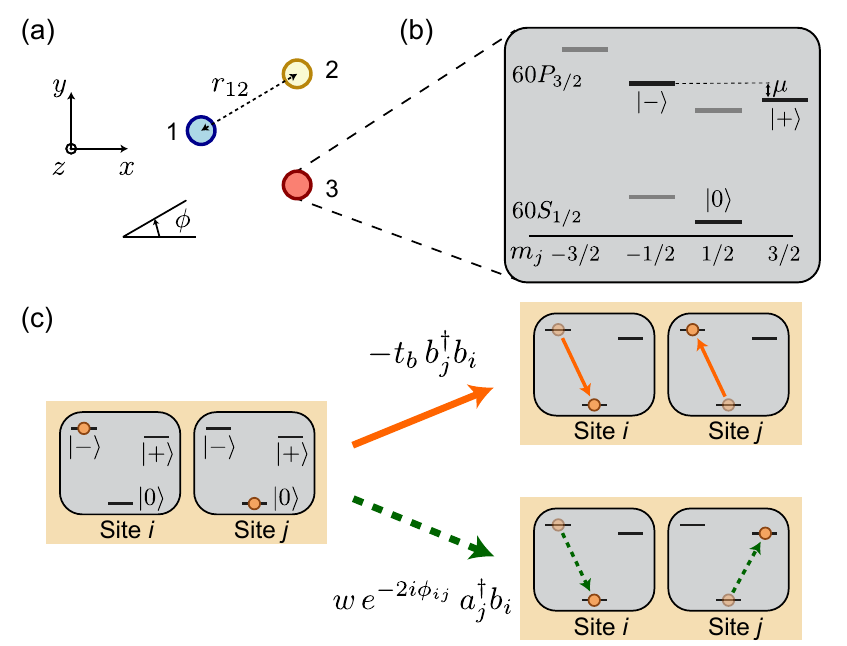}
\caption{{\bf Spin-orbit coupling induced by  dipolar exchange interaction.}
(a) Experimental configuration of three atoms trapped in a tunable geometry. 
The quantization axis $z$, along the magnetic field, is perpendicular to the array of atoms. 
(b) Schematic Zeeman structure of the two Rydberg manifolds $60S_{1/2}$ and $60P_{3/2}$ 
used in this work. The three levels $\ket{0}$, $\ket{+}$ and $\ket{-}$ of the V-structure
involved in the dipole-dipole interaction are indicated
as black lines. The energy difference between $\ket{+}$ and $\ket{-}$ is $\mu$, controlled
by DC magnetic and electric fields perpendicular to the triangle.
(c) The two processes for a $\ket{-}$ excitation to hop from site $i$ to site $j$: 
the $\ket{-}$ excitation is annihilated on site $i$, and a $\ket{-}$ (solid arrow) 
or a $\ket{+}$ (dashed arrow) excitation is created on site $j$.} 
\label{fig:fig1}
\end{figure}

Our system consists of three $^{87}\mathrm{Rb}$ atoms trapped in optical tweezers 
placed in an equilateral configuration, see Fig.~\ref{fig:fig1}(a). 
For each atom, we consider three Rydberg states from the $60S_{1/2}$ and the $60P_{3/2}$ 
manifolds (separated in frequency by 17.2 GHz) in a V-structure, as shown in Fig.~\ref{fig:fig1}(b).  
The state $\ket{0} = \ket{60S_{1/2}, m_j = 1/2 }$ corresponds to the absence of excitation, 
and the two excited states $\ket{+} = \ket{60P_{3/2}, m_j = 3/2 }$ and $\ket{-} = \ket{60P_{3/2}, m_j = -1/2 }$, 
correspond to the two internal states of the excitation. 
We describe these two components of the excitation on a site $i$
by the bosonic operators $a^{\dag}_{i}$ and $b^{\dag}_{i}$ defined by
$a^{\dag}_{i} |0\rangle = |+\rangle_{i}$  and  $b^{\dag}_{i}|0\rangle = |-\rangle_{i}$.
The energy difference $\mu = E_+ - E_-$ between $\ket{+}$ and $\ket{-}$ 
is controlled by a magnetic field $B_z$ and an electric field $E_z$, both orthogonal to the atomic array. 
The excitation transfer between two Rydberg atoms is governed by the dipole-dipole interaction $\hat{V}_{ij}=(\hat{\boldsymbol d}_i\cdot \hat{\boldsymbol d}_j - 3(\hat{\boldsymbol d}_i\cdot \hat{\boldsymbol r})(\hat{\boldsymbol d}_j\cdot \hat{\boldsymbol r}) )/(4 \pi \epsilon_0 r_{ij}^3)$. In our configuration, the unit vector $\hat{\boldsymbol r}=(\cos \phi,\sin\phi,0)$ lies in the $(x,y)$ plane, and $\hat{V}_{ij}$ thus reads
\begin{eqnarray}
\hat{V}_{i j}&= & \frac{1}{4 \pi \epsilon_0 r_{ij}^3}  
\left[ \hat{d}_{i}^{z}\hat{d}_{j}^{z} + \frac{1}{2}\left(\hat{d}^{+}_{i} \hat{d}^{-}_{j} +\hat{d}^{-}_{i} \hat{d}^{+}_{j}
\right)  \right. \label{Eq:dipoleinteraction}\\ & &  
\left.- \frac{3}{2}\left(\hat{d}^{+}_{i} \hat{d}^{+}_{j}  e^{-i 2 \phi_{i j}}+\hat{d}^{-}_{i} \hat{d}^{-}_{j}  e^{i 2 \phi_{i j}}
\right)  \right]. \nonumber 
\end{eqnarray}
Here,  $\hat{d}^{x}_i,\hat{d}^{y}_i,\hat{d}^{z}_i$ are the components of the dipole operator $\hat{\boldsymbol d}_i$, 
$\hat{d}^{\pm}_i= \mp (\hat{d}^{x}_i\pm i \hat{d}^{y}_i)/\sqrt{2}$, and $r_{i j}$ and  $\phi_{ij}$
denote the separation and the polar angle between the two Rydberg atoms. 
The first three terms in Eq.~(\ref{Eq:dipoleinteraction}) correspond to a transfer of excitation 
conserving the total internal angular momentum of the two atoms. 
The last two terms describe the spin-orbit coupling: 
the excitation changes its internal state by two quanta during the transfer, 
and the conservation of the total angular momentum requires 
that the corresponding hopping amplitudes acquire a phase $e^{\pm i  2 \phi_{i j}}$.
Therefore, the dipolar interaction leads to two ways for an excitation 
to hop from site $i$ to site $j$, as illustrated in  Fig.~\ref{fig:fig1}(c):
a resonant process, with amplitude $- t_{a}$ or $-t_{b}$, where the internal state of the excitation is conserved,
and an off-resonant  process (by an energy offset $\mu$) with complex amplitude 
$w\mathrm{e}^{\pm2i\phi_{ij}}$, where the excitation changes its internal state. 
The amplitudes $t_{a,b}$ and $w$ scale as $1/r^3_{ij}$
(see more details in Appendix~\ref{App:simulations}).

\begin{figure}[t!]
\centering
\includegraphics[]{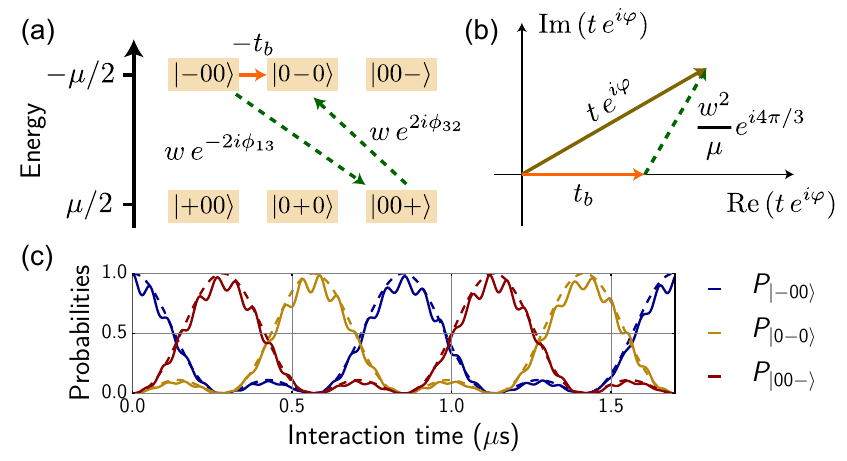}
\caption{{\bf Peierls phase on a triangle.} 
(a) The two available processes for a $\ket{-}$ excitation
to hop from $\ket{-00}$ to $\ket{0 \! - \! 0}$: 
direct hopping with amplitude $-t_b$, or virtual hoppings via $\ket{00+}$. 
(b) Complex plane representation of the effective hopping, 
which is the sum of the two processes depicted in (a). 
(c) Calculated evolution of the site probabilities after preparing $\ket{-00}$ with 
total flux $3\varphi=\pi/2$, for an ideal complex hopping (dashed lines) 
and for our three-level structure involving the $\ket{+}$ states (solid lines). 
The excitation does not spread as time flows, and moves from 
site to site in a chiral way.} 
\label{fig:fig2}
\end{figure}

We now discuss the situation where three atoms are arranged in 
an equilateral triangle and derive the expression of the complex 
hopping amplitude of a $\ket{-}$ excitation. 
We restrict ourselves to the case $\mu\gg t_{a,b}, w$ and treat the hoppings perturbatively. 
As the internal state-flipping hopping is off-resonant, 
the $\ket{-}$ excitation only has a small probability 
of becoming a $\ket{+}$ excitation. 
In addition, as the interaction conserves the number of excitations, 
once the atoms are initialized in the three-site state $\ket{-00}$, 
they mostly remain in the one excitation  subspace consisting of 
the states $\ket{-00}$, $\ket{0 \! - \! 0}$ and $\ket{00-}$.
The hopping of a $\ket{-}$ excitation 
from site 1 to 2, i.e. the change of the three-atom state from
$\ket{-00}$ to $\ket{0\! - \! 0}$ (see Fig~\ref{fig:fig2}(a)), proceeds 
either by a direct hopping with amplitude $-t_{b}$, or by a second-order 
coupling via the intermediate state $\ket{00+}$ consisting in two 
successive flips of the internal state. 
The latter has an amplitude 
$-w^2 \,\mathrm{e}^{2i\left(\phi_{32} - \phi_{13}\right)} / \mu$, with $\phi_{32} - \phi_{13} = 2 \pi/3$.  
Consequently, the hopping amplitude $- t e^{i \varphi}$ from site 1 to  
2 is the sum of the amplitudes of these two processes 
\begin{equation}
t e^{i \varphi} = t_b +  e^{i  4 \pi/3} \frac{w^2}{\mu}.
\label{eq:perturbative}
\end{equation}
The representation of the amplitudes in the complex plane is shown in Figure~\ref{fig:fig2}(b).
In this perturbative picture, the $\ket{+}$ excitation is adiabatically eliminated, 
and the problem reduces to the hopping of the $\ket{-}=b^{\dag}_{i}|0\rangle$ 
excitation described by the effective Hamiltonian  
\begin{equation}\label{Eq:Heff}
H_{\rs eff} = -t \sum_{i =1}^3  \left[e^{i \varphi } b^{\dag}_{i+1}b_{i}+e^{- i \varphi } b^{\dag}_{i} b_{i+1}\right]\ ,
\end{equation}
with  $b_{4}\equiv b_1$. 
The Peierls phase $\varphi$ can be interpreted as the result of an emergent gauge field and  
the magnetic flux through the triangle is thus $3 \varphi$. 
Experimentally, both the effective hopping amplitude $t$ and the flux $3\varphi$ are controlled by the distance between the 
atoms and the energy separation $\mu$. 
For non-zero  flux (modulo $\pi$), 
the excitation exhibits a chiral motion when evolving in the triangle. 
In particular, for $3\varphi= \pm \pi/2$~\cite{Roushan_2016}, 
the excitation hops sequentially from site to site in a preferred direction. 
Figure~\ref{fig:fig2}(c) shows this expected motion 
for the parameters used in the experiment (see Sec.~\ref{sec:chiral_single}): 
we plot the site probabilities as a function of time 
in the case of the complex hopping of a $\ket{-}$ excitation 
described by the Hamiltonian~(\ref{Eq:Heff}) (dashed lines), as well as for
the three-level structure involving the $\ket{+}$ state, 
governed by the Hamiltonian~(\ref{Eq:HeffV}) (solid lines). 
The fast oscillations exhibit a frequency close to $\mu/h$, 
and result from the non-perfect elimination of the $\ket{+}$ state. 

\section{Experimental observation of chiral motion}\label{sec:chiral_single}

\begin{figure*}[t!]
\centering
\includegraphics[]{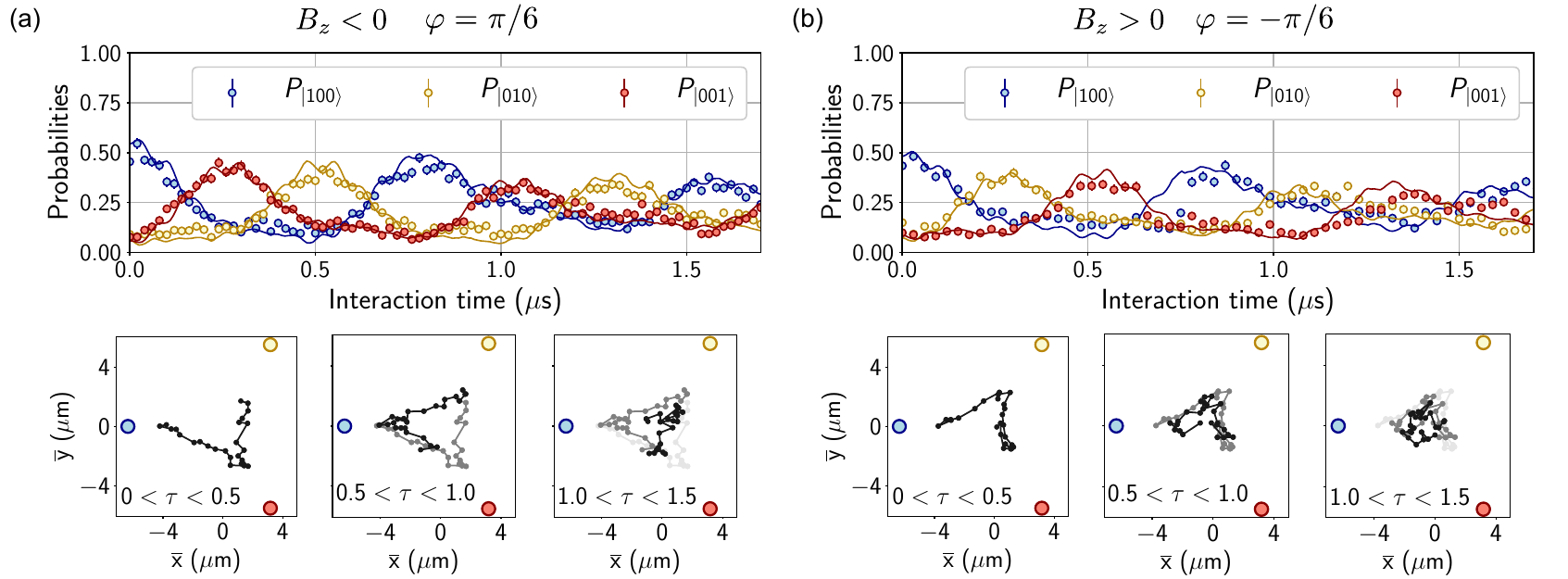}
\caption{{\bf Observation of the chiral motion of a single $\ket{-}$ excitation.}
(a) and (b) Evolution of the three-site probabilities to be in the states  
$\ket{100}$, $\ket{010}$ and $\ket{001}$ as a function of the interaction time 
for two opposite directions of $B_z$. 
Upper panel: experimental results and theoretical 
predictions (solid lines) including experimental errors in the 
preparation and the detection, as well as shot-to-shot fluctuations in the atomic position (which lead to the observed damping of the oscillations). 
Bottom panel: associated trajectories of the center-of-mass 
of the excitation $(\bar{x},\bar{y})$ for specific windows of the excitation time $\tau$, defined by 
$\bar{x}=\sum_{i=1}^3 x_i p_i/\sum_{i=1}^3 p_i$, and 
$\bar{y}=\sum_{i=1}^3 y_i p_i/\sum_{i=1}^3 p_i$ (where $(x_i,y_i)$ 
are the coordinates of site $i$, and $p_i$ the probability for the $\ket{-}$ excitation to be on site $i$).
Error bars denote the standard error on the mean, and are often
smaller than the symbol size.}
\label{fig:fig3}
\end{figure*}

To experimentally demonstrate  the chiral motion of a 
$\ket{-}$ excitation resulting from the complex hopping of Eq.(\ref{eq:perturbative}),
we start with three $^{87}$Rb atoms trapped in 852-nm optical tweezers 
arranged in an equilateral triangle with side length 
$11\, \mu\mathrm{m}$~\cite{Barredo_2016}.  
We optically pump the atoms in the state 
$\ket{5S_{1/2}, F = 2, m_y = -2}$ in $200\, \mu\mathrm{s}$ 
using a quantization axis defined by a magnetic field $B_y>0$ along the $y$-axis 
contained in the triangle plane. 
To isolate the V-structure in the Rydberg manifold 
and achieve isotropic exchange terms $t_{a,b}$ and $w$, 
we must apply static magnetic and electric fields 
perpendicular to the plane of the triangle. 
To do so, we switch on adiabatically $B_z$ and turn off $B_y$ after the 
optical pumping step, in 20~ms: for $B_z<0$ the resulting atomic state 
is thus $\ket{5S_{1/2}, F = 2, m_z = 2}$. 
In order to fix  the Peierls phase to the value leading to the 
chiral motion ($3\varphi \approx \pi/2$), we set $B_z=-8.5$~G and the 
electric field $E_z =0.4$~V/cm, yielding $\mu / h = -16\, \mathrm{MHz}$.
With these values and $r_{ij} = 11\, \mu\mathrm{m}$, we measured, from a
spin exchange experiment with two atoms~\cite{Barredo_2015}, 
$t_{a} / h \simeq 1.5\,\mathrm{MHz}$ and $t_{b} / h \simeq 0.55\,\mathrm{MHz}$ in good agreement
with theoretical calculations of the interaction energies \cite{Weber_2017}.
We then deduce $w / h \simeq 2.7\,\mathrm{MHz}$ using the 
values of the angular part of the dipole matrix elements.
After switching off the dipole traps, we prepare the $\ket{000}$ 
state in $2\,\mu\mathrm{s}$ using a 
STImulated Raman Adiabatic Passage (STIRAP)~\cite{deLeseleuc_2019}, 
via the intermediate state $\ket{5P_{1/2}, F = 2, m_z = 2}$. 
Finally, we address atom 1 with a focused laser 
beam tuned near the $6P_{3/2}-60S_{1/2}$ transition~\cite{deLeseleuc_2017}  
and apply a 400 ns $\pi$-pulse with a microwave resonant with the {\it light-shifted} 
$\ket{0} \rightarrow \ket{-}$ transition. This prepares a $\ket{-}$ excitation on site 1. 

After the preparation of the system in the state $\ket{-00}$, 
we let it evolve under the action of the dipole-dipole interaction for a time $\tau$. 
We then apply a 400 ns read-out pulse to
de-excite the atoms in $\ket{0}$ back to the $5S_{1/2}$ manifold, 
and switch on the dipole traps again. Atoms in the $5S_{1/2}$ state 
are recaptured, whereas atoms still in Rydberg states are lost. 
A final fluorescence image reveals, for each site, 
if the atom is in the $\ket{0}$ state (the atom is recaptured), 
or in another Rydberg state (the atom is lost). 
Our detection method does not distinguish between 
these other Rydberg states, including $\ket{+}$ and $\ket{-}$. 
We will denote the Rydberg states other than $\ket{0}$  as a single state $\ket{1}$. 
As the $\ket{+}$ subspace is hardly 
populated in our experiment, the loss of an atom 
corresponds mainly to its detection in the $\ket{-}$ state. 

The result of this first experiment is presented in Fig.~\ref{fig:fig3}(a), 
where we plot the three-site probabilities to be 
in the states $\ket{100}$, $\ket{010}$ and $\ket{001}$ 
as a function of the interaction time $\tau$. 
As expected, we observe a chiral motion of a localized $\ket{-}$ excitation in the 
counterclockwise direction $1\rightarrow 3\rightarrow 2 \rightarrow 1$. This is the  
signature of an effective magnetic field acting on the hopping excitation, 
described by the Peierls phases. 
The fact that the three probabilities do not sum to 1 
comes from the imperfect preparation of the state $\ket{100}$
 and detection errors.

To reverse the direction of motion, we reverse the sign of $B_z$ 
after the optical pumping stage. The initial atomic state  is now
$\ket{5S_{1/2}, F = 2, m_z = -2}$. 
In this configuration, following the Rydberg excitation,  
the V-structure in the Rydberg manifold involves
$\ket{0}=\ket{60S_{1/2}, m_J = -1/2}$,
$\ket{+}=\ket{60P_{3/2}, m_J = -3/2}$,
and  $\ket{-}=\ket{60P_{3/2}, m_J = +1/2}$. 
The value of  $\mu$ remains unchanged, as the Stark shift only 
depends on $|m_j|$. 
The hopping of a $\ket{-}$ to a $\ket{+}$ excitation now corresponds 
to a {\it decrease}  of the internal momentum by two quanta: 
the orbital phase factor is thus $\mathrm{e}^{2i\phi_{ij}}$, 
and the sign of the Peierls phase is changed. 
Figure~\ref{fig:fig3}(c) shows the same three-site probabilities 
as in Fig.~\ref{fig:fig3}(b) for this opposite direction of $B_z$. 
As expected, we now observe a chiral motion of the $\ket{-}$ 
excitation in the clockwise direction $1\rightarrow 2\rightarrow 3 \rightarrow 1$. 

Finally, we compare the experimental data for the chiral motion in both directions 
with a theoretical model solving the Schr\"odinger equation for this three-atom 
system  including all the Zeeman sublevels of the $60S_{1/2}$ 
and $60P_{3/2}$ manifold. In these simulations, the preparation and detection errors are included
as well as shot-to-shot fluctuations in the atomic positions; the latter leads to small modifications
of the coupling parameters for each shot. The details of these simulations are presented
in Appendix~\ref{App:simulations}. The results are plotted as solid lines on the data in Fig.~\ref{fig:fig3}(a) and (b). 
In both situations, we obtain a good agreement with the model, which reproduces
 the frequency, the amplitude  and the damping of the chiral motion. 

\section{Density-dependent Peierls phase and mapping to anyons}\label{Sec:anyons}

\begin{figure}[t!]
\centering
\includegraphics[]{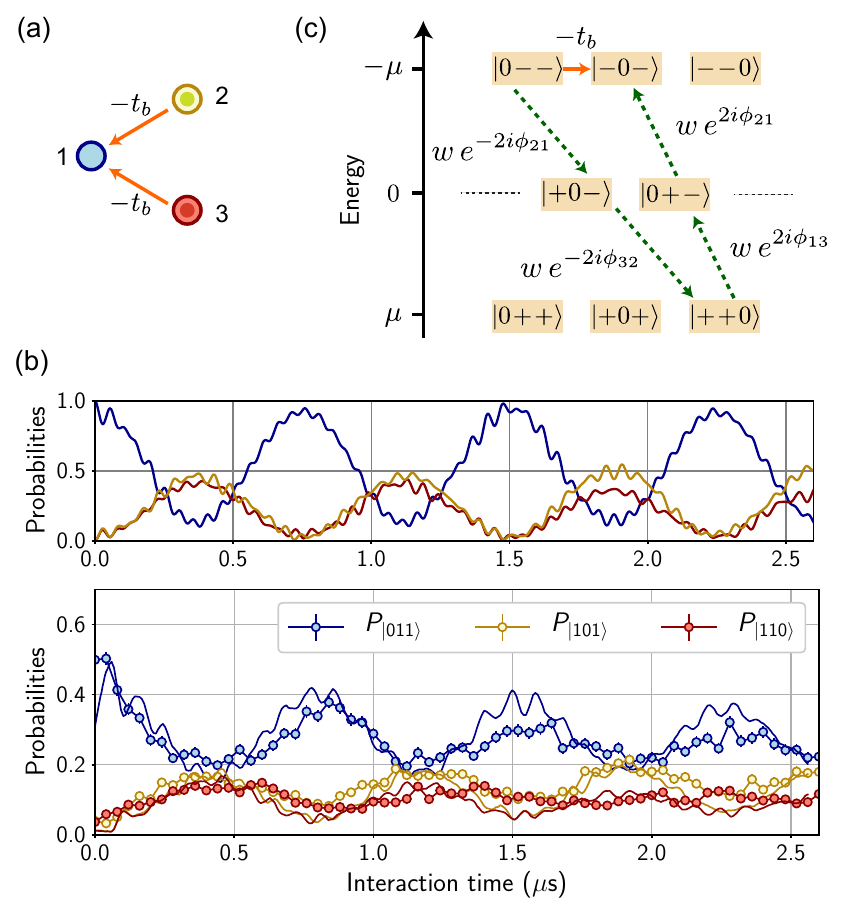}
\caption{{\bf Demonstration of density-dependent hopping for two excitations.}
(a) The presence of a $\ket{-}$ excitation on site 3 
prevents the internal state-flipping process responsible for the complex hopping 
of the $\ket{-}$ excitation from 2 to 1: only the real coupling remains. 
(b) Probability to be in the doubly excited three-site states $\ket{011}$
(targeted initial state), $\ket{101}$ or $\ket{110}$ as a function of the interaction time $\tau$. 
Upper panel: simulations in an ideal case including the three levels of the V-structure. 
Lower panel: experimental results together with the simulation taking into account experimental parameters, including state preparation.
(c) Hopping processes to go from site 1 to site 2 in the two-excitation case, showing
the direct coupling and the  fourth-order process  via $\ket{0 \! + \! +}$. } 
\label{fig:fig4}
\end{figure}

For ensembles of two-level atoms in resonant interaction, 
the excitations can be mapped onto hard-core bosons, 
a fact used in our previous work~\cite{deLeseleuc_2019}. 
A natural question to ask in our present multi-level situation 
is the consequence of the hard-core constraint
on the dynamics of the $\ket{-}$ excitations.
In order to explore this experimentally, 
we now initialize the three-atom system with two $\ket{-}$ excitations on sites 2 and 3, 
while site 1 is in state $\ket{0}$, thus preparing the three atom state $\ket{0 \! - \! -}$. 
To do so we again use the addressing laser on site 1, but tune the $\pi$ microwave 
pulse on resonance with the {\it free space} $\ket{0} \rightarrow \ket{-}$ transition.

In the case of hard-core bosons evolving with the Hamiltonian in Eq.~(\ref{Eq:Heff}),  
one would expect the hole (state $\ket{0}$) 
to propagate in the opposite direction  to the single $\ket{-}$ excitation case, 
as observed using superconducting circuits~\cite{Roushan_2016}. 
The result of our experiment is presented in Fig.~\ref{fig:fig4}, 
where we use the same parameters as for the 
single excitation experiment, i.e., a Peierls phase $\varphi=\pi/6$. 
Remarkably, here we do not observe any chiral motion: 
the hole state $\ket{0}$ propagates almost symmetrically towards sites 2 and 3, 
suggesting that the hopping amplitude between sites is now real, 
and that the description of the dynamics by the Hamiltonian~(\ref{Eq:Heff}) is no longer valid. 
This indicates that the hard-core constraint between the excitations $\ket{-}$ influences the
induced Peierls phases.

To understand this, we come back to the hard-core constraint in our system. 
Two particles, irrespective of their internal state $\ket{+}$ or $\ket{-}$, can not reside on the same site. 
As a consequence, the effective hopping from site 1 to 2 is modified if an excitation 
is already present on site 3: this suppresses the off-resonant process, 
which is at the origin of the complex hopping amplitude in the single excitation case, 
leaving only the direct hopping described by $-t_b$. 
Therefore, the hard-core constraint generates a density-dependent hopping, 
where the phase of the hopping amplitude, as well as its strength, depends 
on the occupation of the third lattice site. 
The effective Hamiltonian describing this situation
generalizes the one of Eq.~(\ref{Eq:Heff}) to 
the case of more than  one $\ket{-}$ excitation:
\begin{equation}\label{Eq:Hanyon}
H_{\rs eff}^{\rs many} = - t \sum_{i=1 }^{3}  \left[e^{i \varphi \left(1-n_{i+2}\right)}b^{\dag}_{i+1} b_{i}+ 
\Delta  b^{\dag}_{i+1} b_{i}  n_{i+2}   + {\rm h.c.}\right] 
\end{equation}
with $n_{i+2} = b^{\dag}_{i+2} b_{i+2}$ the occupation of the third site and $\Delta = (t_{b}-t)/t$. 
The first term in the effective Hamiltonian shows that the Peierls phase is now density-dependent.
The second term describes a conventional correlated hopping, which does not modify
the real or complex nature of the couplings between sites (see Appendix~\ref{App:anyons}). 
In addition, the adiabatic elimination leads 
to two-body interactions terms $\propto (w^2/\mu)n_i n_j$, that do not play a role
in an equilateral triangle and that we therefore drop.

The influence of the density-dependent Peierls phases on the hopping amplitudes
has a simple interpretation in terms of abelian anyonic particles 
in one-dimension in the absence of a magnetic field \cite{Fradkin91,Zhu96,kundu_1999,Keilmann2011,Greschner2015}. 
Here, we obtain anyonic particles with a hard-core constraint and a statistical angle $3 \varphi$.
For this mapping, we use a particle-hole 
transformation and interpret a single hole as an anyonic particle. 
In the absence of a gauge field, a single anyon  (a hole)
exhibits a symmetric dynamics in a triangle, 
which is the result observed in Fig.~\ref{fig:fig4}. 
Now placing two anyons (two holes) in the triangle, 
we are back to the case studied in Sec.~\ref{sec:chiral_single}, where we 
observe a chiral motion (Figure~\ref{fig:fig3}): in the anyon interpretation, 
this is due to the statistical phase under exchange of the two anyonic particles
or equivalently to the fact that one of the two anyonic particles 
carries a magnetic flux for the other one. 
The value of this magnetic flux through the triangle 
is the statistical phase of these anyons. 
The mapping onto anyons can be made rigorous 
and is presented in Appendix~\ref{App:anyons}.

We still observe a residual 
asymmetry in the dynamics, see Fig.~\ref{fig:fig4}(b),
which is also present in the  simulation. 
This indicates that the complex-valued hopping is not fully suppressed. 
Following the same effective Hamiltonian approach 
as the one outlined in Sec.~\ref{sec:spinorbit}, 
the internal state-flipping hopping is now a fourth-order process, as shown in 
Figure~\ref{fig:fig4}(c). 
Considering the hopping from site 1 to site 2, 
the hole can directly hop with an amplitude $- t_b$, 
or virtually go through $\ket{+ \! + \! 0}$, leading to a total amplitude
$te^{i \varphi} = t_b + w^4/\mu^3 e^{-4i\pi/3}$. 
As $w \ll \mu$, the complex part of this hopping is extremely small 
compared to the single particle case, 
thus leading to the observed quasi-symmetric dynamics.

\section{Tunability of the Peierls phase}

\begin{figure}[t!]
\centering
\includegraphics[]{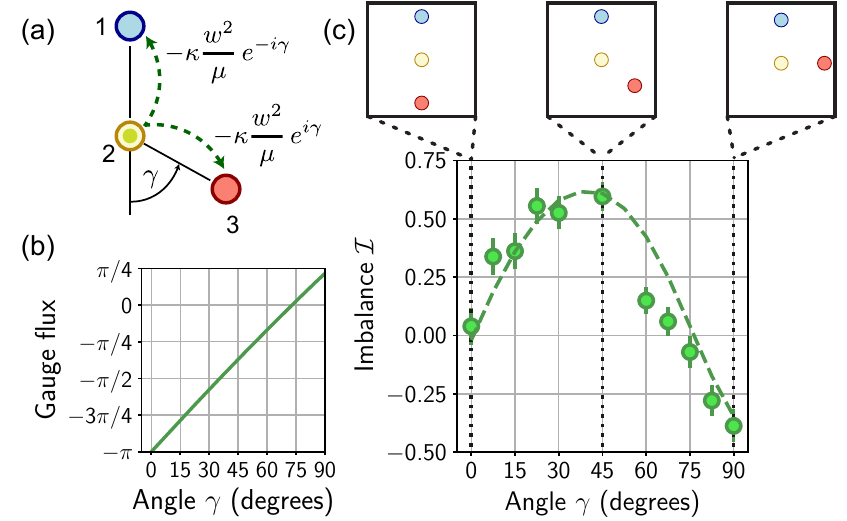}
\caption{{\bf Tunability of the Peierls phase}.
(a) Tunable geometry used for this experiment based on  
an isosceles triangle with $r_{12}=r_{23}=11\,\mu\mathrm{m}$.
(b) Calculated evolution of the magnetic flux threading through 
the isosceles triangle as a function of $\gamma$.
(c) Experimental imbalance  $\cal{I}$ between site 1 and site 3 (see text) after 
having prepared an excitation on site 2 and letting the system evolve 
for $\tau = 0.4\,\mu\mathrm{s}$, as a function of the angle $\gamma$. 
A positive imbalance means that the excitation mainly resides on site 1. 
The three insets represent the triangle configurations 
for three values of $\gamma$, marked on the graph 
by the three dotted lines. 
The dashed line is the simulation.}\label{fig:fig5}
\end{figure}

In a final experiment, we demonstrate the control of the 
Peierls phase in the single excitation case by tuning the 
geometry of the triangle while keeping the same value for $\mu$. 
To do so, we study an isosceles triangle parametrized by the 
angle $\gamma$, see Fig.~\ref{fig:fig5}(a). 
In this configuration, the distance between sites 1 and 3 varies with $\gamma$. 
The effective coupling, and hence the Peierls phase, is then different for each link:
the direct hoppings are 
$t_{12}=t_{23}$ and $t_{13}=\kappa t_{12}$ with $\kappa=1/(2\cos[\gamma/2])^3$; 
the virtual coupling are $\kappa w^2 e^{i\gamma}/\mu $ 
for the $1\rightarrow 2$ and $2\rightarrow 3$ couplings  
and $w^2e^{-2i \gamma}/\mu $ for the $3\rightarrow 1$ coupling. 
The variation of the magnetic flux through the triangle, 
which is the sum of the three Peierls phases, 
is represented in Fig.~\ref{fig:fig5}(b) as a function of the angle $\gamma$. 
It exhibits an almost linear dependence for $\gamma \in \left[0^\circ,90^\circ\right]$.

Our demonstration of the control over the Peierls phase 
is achieved by observing how a single $\ket{-}$ 
excitation prepared initially on site 2 splits  
between site 1 and site 3 after a given evolution time: 
for a negative flux (modulo $2\pi$) the excitation 
propagates towards site 1, while it propagates towards site 3 for a positive flux. 
For zero flux (modulo $\pi$) the propagation is symmetric. 
Fig.~\ref{fig:fig5}(c) shows the result of the experiment. 
We plot the population imbalance between site 1 and site 3, 
${\cal I}=(P_{\ket{100}}-P_{\ket{001}})/(P_{\ket{001}}+P_{\ket{100}})$, 
at time $\tau = 0.4\,\mu\mathrm{s}$, 
as a function of the angle $\gamma$. 
We chose $\tau = 0.4\,\mu\mathrm{s}$ as it corresponds to the excitation 
mainly located on sites 1 and 3 for $\gamma=0^\circ$. 
As expected, we observe that the imbalance varies with the angle $\gamma$, 
and hence with the magnetic flux (Fig.~\ref{fig:fig5}b). 
For $\gamma=0^\circ$ and $75^\circ$ (zero flux) the propagation is symmetric. 
The data are in good agreement with the simulation of the dynamics of the system (dashed line).

\section{Extension to other geometries}

\begin{figure}[t!]
\centering
\includegraphics[]{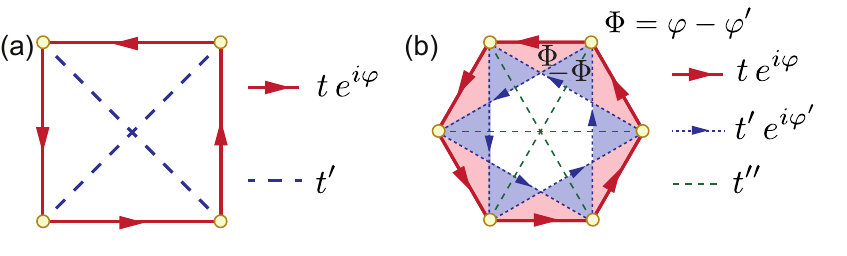}
\caption{{\bf Flux pattern resulting from the complex 
hopping for plaquettes of various geometries.}
(a) Square geometry. The effective Hamiltonian approach yields
$t e^{i\varphi}= t_b + i w^2/(\mu \sqrt{2})$ and $t'=t_b/2^{3/2}-2 w^2/\mu$.
In this case, the  flux  $4 \varphi$ through the square corresponds to a homogeneous 
magnetic field.
(b) Honeycomb geometry. Here $t e^{i\varphi}=t_b +  3w^2/(4\sqrt{3}\mu)e^{i\pi/3}$, 
$t'e^{i\varphi'}=t_b/3^{3/2}+139 w^2/(108\mu)e^{2i\pi/3}$ and 
$t''=t_b/8-4 w^2/(3\sqrt{3}\mu)$.
The flux pattern is well described as an homogeneous magnetic 
field with flux $6 \varphi$ through the honeycomb in combination with
an alternating flux $\Phi=\varphi - \varphi'$ through the red and blue triangles.
}\label{fig:fig6}
\end{figure}

As demonstrated in the previous section for the case of an isosceles triangle, 
the Peierls phase depends on the geometrical arrangement of the atoms.
A natural question to ask is what happens for 
geometries other than a triangle. 
In the following, we discuss theoretically the  
Peierls phase patterns for plaquettes  of square and honeycomb lattices, 
considering the perturbative regime  where the $|+\rangle$ excitation can be eliminated.

For a square geometry, see Fig.~\ref{fig:fig6}(a),  we find a nearest 
neighbor hopping $ t e^{i \varphi}$ with a Peierls phase $\varphi$, as 
the adiabatic elimination gives rise to two distinct virtual 
processes of equal strength. On the contrary, the next nearest neighbor 
hopping remains real valued. Consequently, 
a single excitation experiences a homogeneous  gauge field with a 
flux $4\varphi$ through the square. 
As for the triangle case, the presence of a second excitation gives rise to a 
density-dependent hopping and quenches the virtual processes. 
Therefore,  the dynamics of two excitations is accounted for 
by a modified homogenous magnetic gauge field. 
Finally, for three excitations, all virtual processes are forbidden 
and we recover a time reversal symmetric dynamics.  

For atoms on a honeycomb array, the situation can no longer be described 
by an homogeneous magnetic field. As shown in Fig.~\ref{fig:fig6}(b), 
the Peierls phase $\varphi$ resulting from the nearest neighbor hopping gives 
rise to a homogenous magnetic field with total flux $6 \varphi$. In addition considering 
the next-nearest neighbor coupling introduces a second Peierls phase $\varphi'$. 
The combination of the two phases leads to an alternating flux pattern.  
Such a pattern has been previously 
discussed in connection to the Haldane model on a honeycomb lattice \cite{Haldane1988} 
and provides an intuitive explanation for the appearance of non-trivial 
Chern numbers with $C=\pm 1$ reported in Refs.~\cite{Peter_2015,Weber_2018}. 
For a lattice geometry consisting of many plaquettes, 
the Peierls phase $\varphi$ for the nearest neighbor hopping would now vanish by symmetry, 
whereas the second Peierls phase $\varphi'$ of next nearest hopping remains finite. 
We would thus be able 
to observe chiral edge states in the single-particle regime. 
In Ref.~\cite{Weber_2018} we obtained
these chiral edge states by analyzing the 
band structure of the system and computing the associated Chern numbers 
for the V-structure levels scheme.
The perturbative approach presented here provides  more intuition
on the link between the honeycomb configuration and the Haldane model.

\section{Conclusion}

We have experimentally demonstrated the spin-orbit coupling naturally present in dipolar 
exchange interactions by observing the characteristic chiral motion of an excitation in a minimal setup of 
three Rydberg atoms. A simple explanation of this chiral motion is  achieved in the perturbative regime,
where the spin-orbit coupling gives rise to Peierls phases describing a homogenous magnetic field 
through the triangle. Notably, the Peierls phase depends on the occupation of neighboring sites and
therefore naturally gives rise to a dynamical gauge field. Especially, we have demonstrated in our minimal setup
that this density-dependent Peierls phases can be interpreted as particles with an anyonic exchange statistics.
This minimal setup can be extended to one-dimensional anyon-Hubbard and lattice gauge field models, which will be discussed elsewhere~\cite{InPrep}. By varying the spatial arrangement, we engineered geometry-dependent Peierls phases 
and explored theoretically configurations  
beyond the triangle. In particular for the honeycomb plaquette, 
we showed that at the single-particle level and  in the perturbative approach, 
our system shows the same couplings as those of the
celebrated Haldane model, which is characterized by a non-trivial topological band structure. 
This leads to an intriguing open question,
whether the combination of such topological band structures with the 
strong interactions between the bosonic particles
can lead to the experimental observation of integer or fractional Chern insulators 
\cite{Wan11,Wan12}. 

\begin{acknowledgements}
We thank Hannah Williams for discussions and for her careful reading of the manuscript.
This work benefited from financial support by the EU (FET-Flag 817482, PASQUANS), 
by the R\'egion \^Ile-de-France in the framework of DIM SIRTEQ (project CARAQUES), 
by the IXCORE-Fondation pour la Recherche as well as the French-German collaboration 
for joint projects in NLE Sciences funded by the Deutsche Forschungsgemeinschaft (DFG)
and the Agence National de la Recherche (ANR, project RYBOTIN). M.F. is supported by the Deutsche Forschungsgemeinschaft (DFG) through SFB TR185, project number 277625399.
H.P.B. is supported by the European Union under the ERC consolidator grant SIRPOL (grant no. 681208).  H.P.B., A.B. and M.F. thank the KITP for hospitality.
This research was also supported in part by the National Science Foundation 
under Grant No. NSF PHY-1748958.
\end{acknowledgements}

\appendix

\section{Numerical simulation of the dynamics}\label{App:simulations}

Here, we present numerical simulations of the dynamics of the excitations 
in the triangle, including all the 
Zeeman sublevels of the $60S_{1/2}$ and $60P_{3/2}$ manifolds. 

\begin{figure}[t!]
\centering
\includegraphics{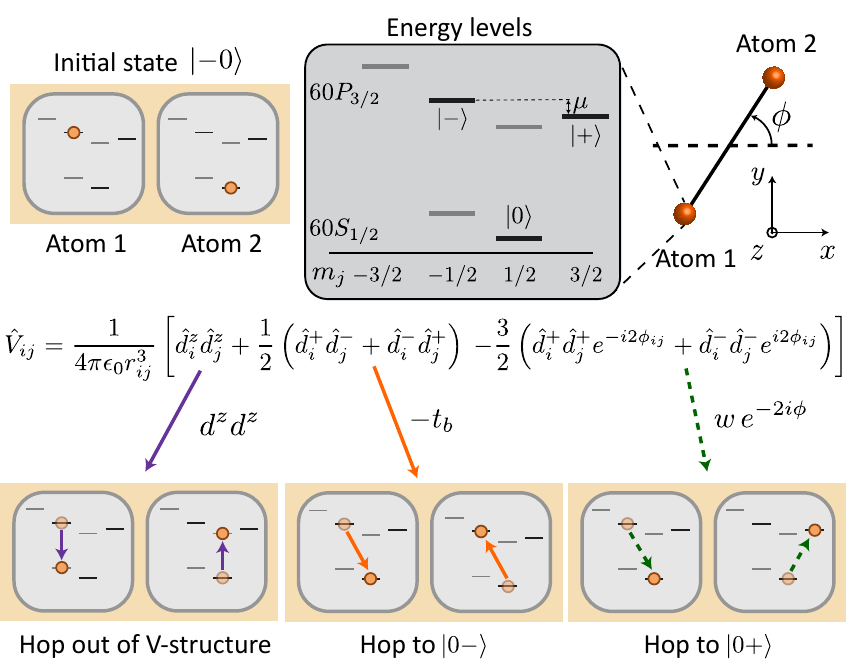}
\caption{{\bf Rydberg levels in a V-structure and hopping processes.}
We consider two atoms as shown on top of the figure, with their six Zeeman 
sub-levels and focus on the V-structure highlighted in black. 
Starting from the initial state $\ket{-0}$, the dipole-dipole interaction in the case of a 
quantization axis perpendicular to the atom array induces three types of hopping. 
The first term of the dipole-dipole interaction makes the system leave the V-structure. 
The two other terms are the direct (solid arrow) and the complex 
(dashed arrow) hoppings, mentioned in the main text.}
\label{fig:S1}
\end{figure}

We first describe the role of the different exchange terms of the dipole-dipole 
interaction of Eq.~(\ref{Eq:dipoleinteraction}) on the various Zeeman states (Figure~\ref{fig:S1}). 
The terms $\hat{d}_i^+\hat{d}_j^-$, $\hat{d}_i^-\hat{d}_j^+$, $\hat{d}_i^+\hat{d}_j^+$ and $\hat{d}_i^-\hat{d}_j^-$ 
keep the system inside the V-structure consisting of the three states $\{\ket{0},\ket{+},\ket{-}\}$.
On the contrary, the $\hat{d}_i^z \hat{d}_j^z$-term couples Zeeman states outside the V-structure. 
The effect of this last term is however inhibited thanks to the electric and magnetic fields, which
energetically isolate the V-structure. In this case, 
the hopping dynamics is described by the Hamiltonian 
\begin{eqnarray}
H&=& \sum_{i\neq j} \left(
a^{\dag}_{i}\ b^{\dag}_{i}\right)  \left(
\begin{array}{c c}
-t_a & w e^{-i 2 \phi_{i j}} \\ 
w e^{i 2 \phi_{i j}} & -t_b
\end{array}
\right) \left(
\begin{array}{c}
a_{j} \\
b_{j}
\end{array}\right) \label{Eq:HeffV} \\
& &+ \sum_{i} \frac{\mu}{2} \left(n_i^{a} - n_{i}^b\right) + H_{\rm vdW},  \nonumber
\end{eqnarray}
where the two bosonic operators $a^{\dag}_{i}$ and $b^{\dag}_{i}$ on site $i$ are defined by  
$a^{\dag}_{i} |0\rangle = |+\rangle_{i}$  
and  $b^{\dag}_{i}|0\rangle = |-\rangle_{i}$. 
The hopping amplitudes $t_{a,b}$ are related to the dipole matrix elements by
\begin{equation}
t_{a,b}={|\bra{\pm} \hat{d}_+\ket{0}|^2 \over 8\pi\epsilon_0 r_{ij}^3},
w={3\bra{+} \hat{d}_+\ket{0}\bra{0} \hat{d}_-\ket{-} \over 8\pi\epsilon_0 r_{ij}^3}\ .
\end{equation}
The term $H_{\rm vdW}$ includes
the van der Waals interactions between the Rydberg levels (typically around 70 kHz), 
which are negligible with respect to the
hopping amplitudes. 
However, we do include it for the quantitative comparison between theory 
and experimental results.

To isolate  the V-structure, we apply DC magnetic and electric fields perpendicular to the triangle. 
The magnetic field of $8.5$~G lifts the degeneracy of the Zeeman 
sub-levels of a single atom. However, the pair state $\ket{--}$ is still 
degenerate with $\ket{60P_{3/2},m_j=-3/2,60P_{3/2},m_j=1/2}$. 
To avoid leakage to this state due to the resonant interaction, 
we lift the degeneracy by additionally 
applying an electric field. We choose  $E_z=0.4$~V/cm, for which 
the static dipole moment induced by the electric field is still small. 
Isolating Rydberg manifolds prevents one from using
short interatomic distances where Rydberg levels get intermixed. 
On the other hand, strong interactions and thus, fast dynamics are 
necessary to neglect the decay of the Rydberg levels and the motion of atoms. 
For our experiment, an interatomic distance of $11\,\mu\mathrm{m}$  
is a good tradeoff. All parameters are optimized under the constraint 
that the condition for chiral propagation is fulfilled. 

As seen in the main text, the dynamics of the excitations in the triangle 
can be qualitatively understood 
by considering only the exchange interactions between the levels 
of the V-structure, see~Fig.\ref{fig:fig2}. Since the isolation of the V-structure 
is in practice not perfect, we perform simulations including all Zeeman 
sublevels of the $60S_{1/2}$ and $60P_{3/2}$ manifolds. 
Considering the Rydberg states outside these two manifolds results in  
van der Waals interactions between the atoms, which we include in the simulation. 
This interaction has however a negligible influence on the dynamics for 
the parameters used in the experiment.  
The strengths of both the resonant and van der Waals interactions
are calculated in the presence of the applied 
electric and magnetic fields using our open source calculator~\cite{Weber_2017}.

The simulation starts with a triangle where each atom is in the $\ket{0}$ state. 
As a first step, we simulate the preparation of the $\ket{-}$ excitations using the addressing beam 
inducing a local light-shift of typically 6 MHz, and the microwave pulse. We take 
into account the van der Waals and exchange interactions between the atoms during the preparation. 
After pre-diagonalizing the single-atom Hamiltonians describing 
the interaction of the atoms with the applied static fields, we introduce 
the microwave couplings. The microwave couples the  Stark- and 
Zeeman-shifted states of the $60S_{1/2}$ manifold to the 
$60P_{3/2}$ manifold. As mentionned in Sec.~\ref{sec:chiral_single}, 
we apply a light-shift to one atom and tune the frequency 
of the microwave to be resonant with the transition to the $\ket{-}$ excitation. 
The computations are performed in the rotating frame within 
the rotating-wave approximation. The simulation of the preparation process 
indicates leakage to other states outside the V-structure, on the order of $5\%$.
Presumably, this leakage could be reduced using optimal control. 
As a second step, we simulate the time evolution of the prepared state under the 
influence of the dipolar exchange interaction.

We take into account experimental imperfections by sampling over 
500 different realizations of the initial configuration of the triangles. 
Firstly, we take for the probability for lattice vacancies  
(due to missing atoms or errors in 
the STIRAP process) the measured value 0.17.
Secondly, we  consider shot-to-shot fluctuations 
of the positions of the atoms in their tweezers, which results in varying hopping strengths. 
Importantly, due to these fluctuations, the atoms can also be positioned 
in such a way that the interatomic axis is not exactly perpendicular 
to the quantization axis. In this case, the dipolar interaction 
can change the magnetic quantum number by one, provoking 
additional leakage to states outside the V-structure. 
These experimental imperfections are responsible 
for the observed damping of the dynamics. 
Finally, detection errors are 
included through a Monte Carlo sampling of the numerical 
results~\cite{deLeseleuc_2018, deLeseleuc_2019}. In the simulations
we account for the fact that the detection scheme does not distinguish between  
states other than $\ket{0}$ by computing the  probabilities $P_{\ket{100}}$, $P_{\ket{010}}$
and $P_{\ket{001}}$ as measured in the experiment.
Note that the preparation of a state with 
two excitations is experimentally challenging and prone to additional errors. 
Therefore, we have scaled vertically the theory curve shown in Fig.~\ref{fig:fig4}(b) by a  factor 0.8.

\section{Mapping onto an anyonic problem}\label{App:anyons}

\subsection{Formal mapping}

Here, we demonstrate that the excitations on a triangle, described by a 
Hamiltonian with the density-dependent Peierls phases,  can be understood as a system 
of hard-core abelian anyons with a non-trivial phase under exchange.  
We start from the Hamiltonian~(\ref{Eq:Hanyon})
\begin{equation}
H_{\rs eff}^{\rs many} = - t \sum_{i=1 }^{3}  \left[e^{i \varphi \left(1-n_{i+2}\right)}b^{\dag}_{i+1} b_{i}+ 
\Delta  b^{\dag}_{i+1} b_{i}  n_{i+2}   + {\rm h.c.}\right] 
\end{equation}
with $n_i=b_i^\dag b_i$. The excitations are described by the bosonic creation 
(annihilation) operators $b^{\dag}_i$ ($b_{i}$), respectively, with $[b_i,b_j^\dag]=0$ and 
$\com{b_i}{b_j}=0 $  for $ i \neq j$. The hard core constraint is most conveniently 
accounted for by the anti-commutation relations 
$ \acom{b_i}{b_i^\dag}=1$ and $\acom{b_i}{b_i}=0$.

In order to map the Hamiltonian to abelian anyons, 
we  define the new modes $B_n$ by the  unitary transformation
\begin{subequations}\label{Eq:a}
\begin{align}
B_1^\dag &\equiv e^{-i\varphi (3-n_2-2n_3)}\,b_1\,, \\
B_2^\dag &\equiv e^{-i\varphi (1+2n_1-3n_3)}\,b_2\,, \\
B_3^\dag &\equiv e^{-i\varphi (n_1-1)}\,b_3 \,.
\end{align}
\end{subequations}
Under this transformation, the Hamiltonian~(\ref{Eq:Hanyon}) now takes the simple form
\begin{equation}\label{Eq:H2}
H=-t\sum_{i=1}^3 \left[ B_{i+1}^\dag B_{i} + \Delta B^{\dag}_{i+1} B_{i} (1- B^\dag_{i+2}B_{i+2}) +  {\rm h.c.} \right],
\end{equation}
which does not feature the density-dependent Peierls phases. 
The second term in Eq.~(\ref{Eq:H2}) describes a conventional correlated hopping where 
the (real) couplings  depend on the number of particles in the new modes.
The influence of the Peierls phases is now hidden in the non-trivial
commutation relations of the modes $B_n$, which can be shown to obey
\begin{subequations}\label{Eq:b}
\begin{align}
\acom{B_n}{B_n^\dag}&=1,\\
\acom{B_n}{B_n}&=0,\\
B_n B_m&=e^{3i\varphi \sign(n-m)}\,B_m B_n, \\
B_n^\dag B_m&=e^{-3i\varphi \sign(n-m)}\,B_m B_n^\dag\ 
\end{align}
\end{subequations}
for $n\neq m$. Here, $\sign(x)=1$ for $x\geq 0$ and $\sign(x)=-1$ for $x<0$. 
This is the algebra of abelian anyons in one dimension with statistical angle
$3\varphi$ and infinite, repulsive on-site interaction (a hard-core constraint)
\cite{Fradkin91}. 
For $\varphi=0$
($\varphi=\pi/3$) we recover hard-core bosons (fermions).
However, for the flux $\pi/2$ with $\varphi=\pi/6$ one finds the non-trivial
``semionic'' commutation relations \cite{zee_1990}
\begin{equation}
B_1B_2=-i\,B_2 B_1
\quad\text{and}\quad
B_1^\dag B_2=i\,B_2 B_1^\dag\quad\text{etc.}
\end{equation}
that describe particles ``halfway'' between bosons and fermions.

\subsection{Interpretation of the dynamics on a triangle}\label{subsec:dynamics}

The dynamical behavior of one or two excitations ($b_i^\dag$) observed  in the experiment
can now be interpreted as follows in the anyonic picture ($B_n^\dag$). 
Due to the implicit particle-hole transformation in (\ref{Eq:a}), a single
excitation $b_i^\dag$ corresponds to a triangle occupied by \textit{two} anyons
$B_n^\dag$. The chiral motion of a single excitation due to the magnetic field
in (\ref{Eq:Heff}) therefore maps onto the chiral motion of a hole flanked by two
anyons in \eqref{Eq:H2}. However, in the anyonic picture, the phase that induces
chiral motion is not due to a magnetic field (which is absent in \eqref{Eq:H2}),
but rather is a consequence of the statistical phase collected by two anyons that
exchange places.
Similarly, the symmetric motion in the case of two excitations
$b_i^\dag$ corresponds to the non-chiral dynamics of a \textit{single} anyon
$B_n^\dag$ subject to (\ref{Eq:H2}), i.e., without magnetic field.


\begin{thebibliography}{54}%
\makeatletter
\providecommand \@ifxundefined [1]{%
 \@ifx{#1\undefined}
}%
\providecommand \@ifnum [1]{%
 \ifnum #1\expandafter \@firstoftwo
 \else \expandafter \@secondoftwo
 \fi
}%
\providecommand \@ifx [1]{%
 \ifx #1\expandafter \@firstoftwo
 \else \expandafter \@secondoftwo
 \fi
}%
\providecommand \natexlab [1]{#1}%
\providecommand \enquote  [1]{``#1''}%
\providecommand \bibnamefont  [1]{#1}%
\providecommand \bibfnamefont [1]{#1}%
\providecommand \citenamefont [1]{#1}%
\providecommand \href@noop [0]{\@secondoftwo}%
\providecommand \href [0]{\begingroup \@sanitize@url \@href}%
\providecommand \@href[1]{\@@startlink{#1}\@@href}%
\providecommand \@@href[1]{\endgroup#1\@@endlink}%
\providecommand \@sanitize@url [0]{\catcode `\\12\catcode `\$12\catcode
  `\&12\catcode `\#12\catcode `\^12\catcode `\_12\catcode `\%12\relax}%
\providecommand \@@startlink[1]{}%
\providecommand \@@endlink[0]{}%
\providecommand \url  [0]{\begingroup\@sanitize@url \@url }%
\providecommand \@url [1]{\endgroup\@href {#1}{\urlprefix }}%
\providecommand \urlprefix  [0]{URL }%
\providecommand \Eprint [0]{\href }%
\providecommand \doibase [0]{http://dx.doi.org/}%
\providecommand \selectlanguage [0]{\@gobble}%
\providecommand \bibinfo  [0]{\@secondoftwo}%
\providecommand \bibfield  [0]{\@secondoftwo}%
\providecommand \translation [1]{[#1]}%
\providecommand \BibitemOpen [0]{}%
\providecommand \bibitemStop [0]{}%
\providecommand \bibitemNoStop [0]{.\EOS\space}%
\providecommand \EOS [0]{\spacefactor3000\relax}%
\providecommand \BibitemShut  [1]{\csname bibitem#1\endcsname}%
\let\auto@bib@innerbib\@empty
\bibitem [{\citenamefont {Georgescu}\ \emph {et~al.}(2014)\citenamefont
  {Georgescu}, \citenamefont {Ashhab},\ and\ \citenamefont
  {Nori}}]{Georgescu2014}%
  \BibitemOpen
  \bibfield  {author} {\bibinfo {author} {\bibfnamefont {I.~M.}\ \bibnamefont
  {Georgescu}}, \bibinfo {author} {\bibfnamefont {S.}~\bibnamefont {Ashhab}}, \
  and\ \bibinfo {author} {\bibfnamefont {F.}~\bibnamefont {Nori}},\ }\bibfield
  {title} {\enquote {\bibinfo {title} {{Q}uantum simulation},}\ }\href
  {\doibase 10.1103/RevModPhys.86.153} {\bibfield  {journal} {\bibinfo
  {journal} {Rev. Mod. Phys.}\ }\textbf {\bibinfo {volume} {86}},\ \bibinfo
  {pages} {153} (\bibinfo {year} {2014})}\BibitemShut {NoStop}%
\bibitem [{\citenamefont {Bergholtz}\ and\ \citenamefont
  {Liu}(2013)}]{Bergholtz2013}%
  \BibitemOpen
  \bibfield  {author} {\bibinfo {author} {\bibfnamefont {E.~J.}\ \bibnamefont
  {Bergholtz}}\ and\ \bibinfo {author} {\bibfnamefont {Z.}~\bibnamefont
  {Liu}},\ }\bibfield  {title} {\enquote {\bibinfo {title} {Topological flat
  band models and fractional {C}hern insulators},}\ }\href {\doibase
  10.1142/S021797921330017X} {\bibfield  {journal} {\bibinfo  {journal}
  {International Journal of Modern Physics B}\ }\textbf {\bibinfo {volume}
  {27}},\ \bibinfo {pages} {1330017} (\bibinfo {year} {2013})}\BibitemShut
  {NoStop}%
\bibitem [{\citenamefont {Cooper}\ \emph {et~al.}(2019)\citenamefont {Cooper},
  \citenamefont {Dalibard},\ and\ \citenamefont {Spielman}}]{Cooper2019}%
  \BibitemOpen
  \bibfield  {author} {\bibinfo {author} {\bibfnamefont {N.~R.}\ \bibnamefont
  {Cooper}}, \bibinfo {author} {\bibfnamefont {J.}~\bibnamefont {Dalibard}}, \
  and\ \bibinfo {author} {\bibfnamefont {I.~B.}\ \bibnamefont {Spielman}},\
  }\bibfield  {title} {\enquote {\bibinfo {title} {Topological bands for
  ultracold atoms},}\ }\href {\doibase 10.1103/RevModPhys.91.015005} {\bibfield
   {journal} {\bibinfo  {journal} {Rev. Mod. Phys.}\ }\textbf {\bibinfo
  {volume} {91}},\ \bibinfo {pages} {015005} (\bibinfo {year}
  {2019})}\BibitemShut {NoStop}%
\bibitem [{\citenamefont {Hofstadter}(1976)}]{Hofstadter1976}%
  \BibitemOpen
  \bibfield  {author} {\bibinfo {author} {\bibfnamefont {D.~R.}\ \bibnamefont
  {Hofstadter}},\ }\bibfield  {title} {\enquote {\bibinfo {title} {{Energy
  levels and wave functions of Bloch electrons in rational and irrational
  magnetic fields}},}\ }\href {\doibase 10.1103/PhysRevB.14.2239} {\bibfield
  {journal} {\bibinfo  {journal} {Phys. Rev. B}\ }\textbf {\bibinfo {volume}
  {14}},\ \bibinfo {pages} {2239} (\bibinfo {year} {1976})}\BibitemShut
  {NoStop}%
\bibitem [{\citenamefont {Jaksch}\ and\ \citenamefont
  {Zoller}(2003)}]{Jaksch_2003}%
  \BibitemOpen
  \bibfield  {author} {\bibinfo {author} {\bibfnamefont {D.}~\bibnamefont
  {Jaksch}}\ and\ \bibinfo {author} {\bibfnamefont {P.}~\bibnamefont
  {Zoller}},\ }\bibfield  {title} {\enquote {\bibinfo {title} {{Creation of
  effective magnetic fields in optical lattices: the Hofstadter butterfly for
  cold neutral atoms}},}\ }\href {\doibase 10.1088/1367-2630/5/1/356}
  {\bibfield  {journal} {\bibinfo  {journal} {New Journal of Physics}\ }\textbf
  {\bibinfo {volume} {5}},\ \bibinfo {pages} {56} (\bibinfo {year}
  {2003})}\BibitemShut {NoStop}%
\bibitem [{\citenamefont {Goldman}\ \emph {et~al.}(2014)\citenamefont
  {Goldman}, \citenamefont {Juzeli{\={u}}nas}, \citenamefont {{\"O}hberg},\
  and\ \citenamefont {Spielman}}]{Goldman_2014}%
  \BibitemOpen
  \bibfield  {author} {\bibinfo {author} {\bibfnamefont {N.}~\bibnamefont
  {Goldman}}, \bibinfo {author} {\bibfnamefont {G.}~\bibnamefont
  {Juzeli{\={u}}nas}}, \bibinfo {author} {\bibfnamefont {P.}~\bibnamefont
  {{\"O}hberg}}, \ and\ \bibinfo {author} {\bibfnamefont {I.~B.}\ \bibnamefont
  {Spielman}},\ }\bibfield  {title} {\enquote {\bibinfo {title} {Light-induced
  gauge fields for ultracold atoms},}\ }\href {\doibase
  10.1088/0034-4885/77/12/126401} {\bibfield  {journal} {\bibinfo  {journal}
  {Rep. Prog. Phys.}\ }\textbf {\bibinfo {volume} {77}},\ \bibinfo {pages}
  {126401} (\bibinfo {year} {2014})}\BibitemShut {NoStop}%
\bibitem [{\citenamefont {Dalibard}\ \emph {et~al.}(2011)\citenamefont
  {Dalibard}, \citenamefont {Gerbier}, \citenamefont
  {Juzeli\ifmmode~\bar{u}\else \={u}\fi{}nas},\ and\ \citenamefont
  {\"Ohberg}}]{Dalibard_2010}%
  \BibitemOpen
  \bibfield  {author} {\bibinfo {author} {\bibfnamefont {J.}~\bibnamefont
  {Dalibard}}, \bibinfo {author} {\bibfnamefont {F.}~\bibnamefont {Gerbier}},
  \bibinfo {author} {\bibfnamefont {G.}~\bibnamefont
  {Juzeli\ifmmode~\bar{u}\else \={u}\fi{}nas}}, \ and\ \bibinfo {author}
  {\bibfnamefont {P.}~\bibnamefont {\"Ohberg}},\ }\bibfield  {title} {\enquote
  {\bibinfo {title} {Colloquium: Artificial gauge potentials for neutral
  atoms},}\ }\href {\doibase 10.1103/RevModPhys.83.1523} {\bibfield  {journal}
  {\bibinfo  {journal} {Rev. Mod. Phys.}\ }\textbf {\bibinfo {volume} {83}},\
  \bibinfo {pages} {1523} (\bibinfo {year} {2011})}\BibitemShut {NoStop}%
\bibitem [{\citenamefont {Galitski}\ and\ \citenamefont
  {Spielman}(2013)}]{Spielman_2013}%
  \BibitemOpen
  \bibfield  {author} {\bibinfo {author} {\bibfnamefont {V.}~\bibnamefont
  {Galitski}}\ and\ \bibinfo {author} {\bibfnamefont {I.~B.}\ \bibnamefont
  {Spielman}},\ }\bibfield  {title} {\enquote {\bibinfo {title} {{Spin-orbit
  coupling in quantum gases}},}\ }\href {\doibase 10.1038/nature11841}
  {\bibfield  {journal} {\bibinfo  {journal} {Nature}\ }\textbf {\bibinfo
  {volume} {494}},\ \bibinfo {pages} {49} (\bibinfo {year} {2013})}\BibitemShut
  {NoStop}%
\bibitem [{\citenamefont {Zhai}(2015)}]{Zhai_2015}%
  \BibitemOpen
  \bibfield  {author} {\bibinfo {author} {\bibfnamefont {H.}~\bibnamefont
  {Zhai}},\ }\bibfield  {title} {\enquote {\bibinfo {title} {{Degenerate
  quantum gases with spin{\textendash}orbit coupling: a review}},}\ }\href
  {\doibase 10.1088/0034-4885/78/2/026001} {\bibfield  {journal} {\bibinfo
  {journal} {Reports on Progress in Physics}\ }\textbf {\bibinfo {volume}
  {78}},\ \bibinfo {pages} {026001} (\bibinfo {year} {2015})}\BibitemShut
  {NoStop}%
\bibitem [{\citenamefont {Aidelsburger}\ \emph {et~al.}(2018)\citenamefont
  {Aidelsburger}, \citenamefont {Nascimbene},\ and\ \citenamefont
  {Goldman}}]{Aidelsburger2018}%
  \BibitemOpen
  \bibfield  {author} {\bibinfo {author} {\bibfnamefont {M.}~\bibnamefont
  {Aidelsburger}}, \bibinfo {author} {\bibfnamefont {S.}~\bibnamefont
  {Nascimbene}}, \ and\ \bibinfo {author} {\bibfnamefont {N.}~\bibnamefont
  {Goldman}},\ }\bibfield  {title} {\enquote {\bibinfo {title} {Artificial
  gauge fields in materials and engineered systems},}\ }\href {\doibase
  https://doi.org/10.1016/j.crhy.2018.03.002} {\bibfield  {journal} {\bibinfo
  {journal} {Comptes Rendus Physique}\ }\textbf {\bibinfo {volume} {19}},\
  \bibinfo {pages} {394 } (\bibinfo {year} {2018})}\BibitemShut {NoStop}%
\bibitem [{\citenamefont {Aidelsburger}\ \emph {et~al.}(2011)\citenamefont
  {Aidelsburger}, \citenamefont {Atala}, \citenamefont {Nascimb\`ene},
  \citenamefont {Trotzky}, \citenamefont {Chen},\ and\ \citenamefont
  {Bloch}}]{Aidelsburger_2011}%
  \BibitemOpen
  \bibfield  {author} {\bibinfo {author} {\bibfnamefont {M.}~\bibnamefont
  {Aidelsburger}}, \bibinfo {author} {\bibfnamefont {M.}~\bibnamefont {Atala}},
  \bibinfo {author} {\bibfnamefont {S.}~\bibnamefont {Nascimb\`ene}}, \bibinfo
  {author} {\bibfnamefont {S.}~\bibnamefont {Trotzky}}, \bibinfo {author}
  {\bibfnamefont {Y.-A.}\ \bibnamefont {Chen}}, \ and\ \bibinfo {author}
  {\bibfnamefont {I.}~\bibnamefont {Bloch}},\ }\bibfield  {title} {\enquote
  {\bibinfo {title} {Experimental realization of strong effective magnetic
  fields in an optical lattice},}\ }\href {\doibase
  10.1103/PhysRevLett.107.255301} {\bibfield  {journal} {\bibinfo  {journal}
  {Phys. Rev. Lett.}\ }\textbf {\bibinfo {volume} {107}},\ \bibinfo {pages}
  {255301} (\bibinfo {year} {2011})}\BibitemShut {NoStop}%
\bibitem [{\citenamefont {Kolovsky}(2011)}]{Kolovsky2011}%
  \BibitemOpen
  \bibfield  {author} {\bibinfo {author} {\bibfnamefont {A.~R.}\ \bibnamefont
  {Kolovsky}},\ }\bibfield  {title} {\enquote {\bibinfo {title} {Creating
  artificial magnetic fields for cold atoms by photon-assisted tunneling},}\
  }\href {\doibase 10.1209/0295-5075/93/20003} {\bibfield  {journal} {\bibinfo
  {journal} {{EPL} (Europhysics Letters)}\ }\textbf {\bibinfo {volume} {93}},\
  \bibinfo {pages} {20003} (\bibinfo {year} {2011})}\BibitemShut {NoStop}%
\bibitem [{\citenamefont {Struck}\ \emph {et~al.}(2012)\citenamefont {Struck},
  \citenamefont {\"Olschl\"ager}, \citenamefont {Weinberg}, \citenamefont
  {Hauke}, \citenamefont {Simonet}, \citenamefont {Eckardt}, \citenamefont
  {Lewenstein}, \citenamefont {Sengstock},\ and\ \citenamefont
  {Windpassinger}}]{Struck_2012}%
  \BibitemOpen
  \bibfield  {author} {\bibinfo {author} {\bibfnamefont {J.}~\bibnamefont
  {Struck}}, \bibinfo {author} {\bibfnamefont {C.}~\bibnamefont
  {\"Olschl\"ager}}, \bibinfo {author} {\bibfnamefont {M.}~\bibnamefont
  {Weinberg}}, \bibinfo {author} {\bibfnamefont {P.}~\bibnamefont {Hauke}},
  \bibinfo {author} {\bibfnamefont {J.}~\bibnamefont {Simonet}}, \bibinfo
  {author} {\bibfnamefont {A.}~\bibnamefont {Eckardt}}, \bibinfo {author}
  {\bibfnamefont {M.}~\bibnamefont {Lewenstein}}, \bibinfo {author}
  {\bibfnamefont {K.}~\bibnamefont {Sengstock}}, \ and\ \bibinfo {author}
  {\bibfnamefont {P.}~\bibnamefont {Windpassinger}},\ }\bibfield  {title}
  {\enquote {\bibinfo {title} {Tunable gauge potential for neutral and spinless
  particles in driven optical lattices},}\ }\href {\doibase
  10.1103/PhysRevLett.108.225304} {\bibfield  {journal} {\bibinfo  {journal}
  {Phys. Rev. Lett.}\ }\textbf {\bibinfo {volume} {108}},\ \bibinfo {pages}
  {225304} (\bibinfo {year} {2012})}\BibitemShut {NoStop}%
\bibitem [{\citenamefont {Jotzu}\ \emph {et~al.}(2014)\citenamefont {Jotzu},
  \citenamefont {Messer}, \citenamefont {Desbuquois}, \citenamefont {Lebrat},
  \citenamefont {Uehlinger}, \citenamefont {Greif},\ and\ \citenamefont
  {Esslinger}}]{Jotzu2014}%
  \BibitemOpen
  \bibfield  {author} {\bibinfo {author} {\bibfnamefont {G.}~\bibnamefont
  {Jotzu}}, \bibinfo {author} {\bibfnamefont {M.}~\bibnamefont {Messer}},
  \bibinfo {author} {\bibfnamefont {R.}~\bibnamefont {Desbuquois}}, \bibinfo
  {author} {\bibfnamefont {M.}~\bibnamefont {Lebrat}}, \bibinfo {author}
  {\bibfnamefont {T.}~\bibnamefont {Uehlinger}}, \bibinfo {author}
  {\bibfnamefont {D.}~\bibnamefont {Greif}}, \ and\ \bibinfo {author}
  {\bibfnamefont {T.}~\bibnamefont {Esslinger}},\ }\bibfield  {title} {\enquote
  {\bibinfo {title} {{Experimental realization of the topological Haldane model
  with ultracold fermions}},}\ }\href {\doibase 10.1038/nature13915} {\bibfield
   {journal} {\bibinfo  {journal} {Nature}\ }\textbf {\bibinfo {volume}
  {515}},\ \bibinfo {pages} {237} (\bibinfo {year} {2014})}\BibitemShut
  {NoStop}%
\bibitem [{\citenamefont {Mancini}\ \emph {et~al.}(2015)\citenamefont
  {Mancini}, \citenamefont {Pagano}, \citenamefont {Cappellini}, \citenamefont
  {Livi}, \citenamefont {Rider}, \citenamefont {Catani}, \citenamefont {Sias},
  \citenamefont {Zoller}, \citenamefont {Inguscio}, \citenamefont {Dalmonte},\
  and\ \citenamefont {Fallani}}]{Mancini_2015}%
  \BibitemOpen
  \bibfield  {author} {\bibinfo {author} {\bibfnamefont {M.}~\bibnamefont
  {Mancini}}, \bibinfo {author} {\bibfnamefont {G.}~\bibnamefont {Pagano}},
  \bibinfo {author} {\bibfnamefont {G.}~\bibnamefont {Cappellini}}, \bibinfo
  {author} {\bibfnamefont {L.}~\bibnamefont {Livi}}, \bibinfo {author}
  {\bibfnamefont {M.}~\bibnamefont {Rider}}, \bibinfo {author} {\bibfnamefont
  {J.}~\bibnamefont {Catani}}, \bibinfo {author} {\bibfnamefont
  {C.}~\bibnamefont {Sias}}, \bibinfo {author} {\bibfnamefont {P.}~\bibnamefont
  {Zoller}}, \bibinfo {author} {\bibfnamefont {M.}~\bibnamefont {Inguscio}},
  \bibinfo {author} {\bibfnamefont {M.}~\bibnamefont {Dalmonte}}, \ and\
  \bibinfo {author} {\bibfnamefont {L.}~\bibnamefont {Fallani}},\ }\bibfield
  {title} {\enquote {\bibinfo {title} {{Observation of chiral edge states with
  neutral fermions in synthetic Hall ribbons}},}\ }\href {\doibase
  10.1126/science.aaa8736} {\bibfield  {journal} {\bibinfo  {journal}
  {Science}\ }\textbf {\bibinfo {volume} {349}},\ \bibinfo {pages} {1510}
  (\bibinfo {year} {2015})}\BibitemShut {NoStop}%
\bibitem [{\citenamefont {Stuhl}\ \emph {et~al.}(2015)\citenamefont {Stuhl},
  \citenamefont {Lu}, \citenamefont {Aycock}, \citenamefont {Genkina},\ and\
  \citenamefont {Spielman}}]{Stuhl_2015}%
  \BibitemOpen
  \bibfield  {author} {\bibinfo {author} {\bibfnamefont {B.~K.}\ \bibnamefont
  {Stuhl}}, \bibinfo {author} {\bibfnamefont {H.-I.}\ \bibnamefont {Lu}},
  \bibinfo {author} {\bibfnamefont {L.~M.}\ \bibnamefont {Aycock}}, \bibinfo
  {author} {\bibfnamefont {D.}~\bibnamefont {Genkina}}, \ and\ \bibinfo
  {author} {\bibfnamefont {I.~B.}\ \bibnamefont {Spielman}},\ }\bibfield
  {title} {\enquote {\bibinfo {title} {{Visualizing edge states with an atomic
  Bose gas in the quantum Hall regime}},}\ }\href {\doibase
  10.1126/science.aaa8515} {\bibfield  {journal} {\bibinfo  {journal}
  {Science}\ }\textbf {\bibinfo {volume} {349}},\ \bibinfo {pages} {1514}
  (\bibinfo {year} {2015})}\BibitemShut {NoStop}%
\bibitem [{\citenamefont {Chalopin}\ \emph {et~al.}(2020)\citenamefont
  {Chalopin}, \citenamefont {Satoor}, \citenamefont {Evrard}, \citenamefont
  {Makhalov}, \citenamefont {Dalibard}, \citenamefont {Lopes},\ and\
  \citenamefont {Nascimb\`ene}}]{Nascimbene2020}%
  \BibitemOpen
  \bibfield  {author} {\bibinfo {author} {\bibfnamefont {T.}~\bibnamefont
  {Chalopin}}, \bibinfo {author} {\bibfnamefont {T.}~\bibnamefont {Satoor}},
  \bibinfo {author} {\bibfnamefont {A.}~\bibnamefont {Evrard}}, \bibinfo
  {author} {\bibfnamefont {V.}~\bibnamefont {Makhalov}}, \bibinfo {author}
  {\bibfnamefont {J.}~\bibnamefont {Dalibard}}, \bibinfo {author}
  {\bibfnamefont {R.}~\bibnamefont {Lopes}}, \ and\ \bibinfo {author}
  {\bibfnamefont {S.}~\bibnamefont {Nascimb\`ene}},\ }\bibfield  {title}
  {\enquote {\bibinfo {title} {{Exploring the topology of a quantum Hall system
  at the microscopic level}},}\ }\href {https://arxiv.org/pdf/2001.01664.pdf}
  {\bibfield  {journal} {\bibinfo  {journal} {arXiv:2001.01664}\ } (\bibinfo
  {year} {2020})}\BibitemShut {NoStop}%
\bibitem [{\citenamefont {Roushan}\ \emph {et~al.}(2017)\citenamefont
  {Roushan}, \citenamefont {Neill}, \citenamefont {Megrant}, \citenamefont
  {Chen}, \citenamefont {Babbush}, \citenamefont {Barends}, \citenamefont
  {Campbell}, \citenamefont {Chen}, \citenamefont {Chiaro}, \citenamefont
  {Dunsworth}, \citenamefont {Fowler}, \citenamefont {Jeffrey}, \citenamefont
  {Kelly}, \citenamefont {Lucero}, \citenamefont {Mutus}, \citenamefont
  {O'Malley}, \citenamefont {Neeley}, \citenamefont {Quintana}, \citenamefont
  {Sank}, \citenamefont {Vainsencher}, \citenamefont {Wenner}, \citenamefont
  {White}, \citenamefont {Kapit}, \citenamefont {Neven},\ and\ \citenamefont
  {Martinis}}]{Roushan_2016}%
  \BibitemOpen
  \bibfield  {author} {\bibinfo {author} {\bibfnamefont {P.}~\bibnamefont
  {Roushan}}, \bibinfo {author} {\bibfnamefont {C.}~\bibnamefont {Neill}},
  \bibinfo {author} {\bibfnamefont {A.}~\bibnamefont {Megrant}}, \bibinfo
  {author} {\bibfnamefont {Y.}~\bibnamefont {Chen}}, \bibinfo {author}
  {\bibfnamefont {R.}~\bibnamefont {Babbush}}, \bibinfo {author} {\bibfnamefont
  {R.}~\bibnamefont {Barends}}, \bibinfo {author} {\bibfnamefont
  {B.}~\bibnamefont {Campbell}}, \bibinfo {author} {\bibfnamefont
  {Z.}~\bibnamefont {Chen}}, \bibinfo {author} {\bibfnamefont {B.}~\bibnamefont
  {Chiaro}}, \bibinfo {author} {\bibfnamefont {A.}~\bibnamefont {Dunsworth}},
  \bibinfo {author} {\bibfnamefont {A.}~\bibnamefont {Fowler}}, \bibinfo
  {author} {\bibfnamefont {E.}~\bibnamefont {Jeffrey}}, \bibinfo {author}
  {\bibfnamefont {J.}~\bibnamefont {Kelly}}, \bibinfo {author} {\bibfnamefont
  {E.}~\bibnamefont {Lucero}}, \bibinfo {author} {\bibfnamefont
  {J.}~\bibnamefont {Mutus}}, \bibinfo {author} {\bibfnamefont {P.~J.~J.}\
  \bibnamefont {O'Malley}}, \bibinfo {author} {\bibfnamefont {M.}~\bibnamefont
  {Neeley}}, \bibinfo {author} {\bibfnamefont {C.}~\bibnamefont {Quintana}},
  \bibinfo {author} {\bibfnamefont {D.}~\bibnamefont {Sank}}, \bibinfo {author}
  {\bibfnamefont {A.}~\bibnamefont {Vainsencher}}, \bibinfo {author}
  {\bibfnamefont {J.}~\bibnamefont {Wenner}}, \bibinfo {author} {\bibfnamefont
  {T.}~\bibnamefont {White}}, \bibinfo {author} {\bibfnamefont
  {E.}~\bibnamefont {Kapit}}, \bibinfo {author} {\bibfnamefont
  {H.}~\bibnamefont {Neven}}, \ and\ \bibinfo {author} {\bibfnamefont
  {J.}~\bibnamefont {Martinis}},\ }\bibfield  {title} {\enquote {\bibinfo
  {title} {Chiral ground-state currents of interacting photons in a synthetic
  magnetic field},}\ }\href {\doibase 10.1038/nphys3930} {\bibfield  {journal}
  {\bibinfo  {journal} {Nature Physics}\ }\textbf {\bibinfo {volume} {13}},\
  \bibinfo {pages} {146} (\bibinfo {year} {2017})}\BibitemShut {NoStop}%
\bibitem [{\citenamefont {Ozawa}\ \emph {et~al.}(2019)\citenamefont {Ozawa},
  \citenamefont {Price}, \citenamefont {Amo}, \citenamefont {Goldman},
  \citenamefont {Hafezi}, \citenamefont {Lu}, \citenamefont {Rechtsman},
  \citenamefont {Schuster}, \citenamefont {Simon}, \citenamefont {Zilberberg},\
  and\ \citenamefont {Carusotto}}]{ReviewPhotonics}%
  \BibitemOpen
  \bibfield  {author} {\bibinfo {author} {\bibfnamefont {T.}~\bibnamefont
  {Ozawa}}, \bibinfo {author} {\bibfnamefont {H.~M.}\ \bibnamefont {Price}},
  \bibinfo {author} {\bibfnamefont {A.}~\bibnamefont {Amo}}, \bibinfo {author}
  {\bibfnamefont {N.}~\bibnamefont {Goldman}}, \bibinfo {author} {\bibfnamefont
  {M.}~\bibnamefont {Hafezi}}, \bibinfo {author} {\bibfnamefont
  {L.}~\bibnamefont {Lu}}, \bibinfo {author} {\bibfnamefont {M.~C.}\
  \bibnamefont {Rechtsman}}, \bibinfo {author} {\bibfnamefont {D.}~\bibnamefont
  {Schuster}}, \bibinfo {author} {\bibfnamefont {J.}~\bibnamefont {Simon}},
  \bibinfo {author} {\bibfnamefont {O.}~\bibnamefont {Zilberberg}}, \ and\
  \bibinfo {author} {\bibfnamefont {I.}~\bibnamefont {Carusotto}},\ }\bibfield
  {title} {\enquote {\bibinfo {title} {Topological photonics},}\ }\href
  {\doibase 10.1103/RevModPhys.91.015006} {\bibfield  {journal} {\bibinfo
  {journal} {Rev. Mod. Phys.}\ }\textbf {\bibinfo {volume} {91}},\ \bibinfo
  {pages} {015006} (\bibinfo {year} {2019})}\BibitemShut {NoStop}%
\bibitem [{\citenamefont {Liu}\ \emph {et~al.}()\citenamefont {Liu},
  \citenamefont {Chen},\ and\ \citenamefont {Xu}}]{ReviewPhononics}%
  \BibitemOpen
  \bibfield  {author} {\bibinfo {author} {\bibfnamefont {Y.}~\bibnamefont
  {Liu}}, \bibinfo {author} {\bibfnamefont {X.}~\bibnamefont {Chen}}, \ and\
  \bibinfo {author} {\bibfnamefont {Y.}~\bibnamefont {Xu}},\ }\bibfield
  {title} {\enquote {\bibinfo {title} {Topological phononics: From fundamental
  models to real materials},}\ }\href {\doibase 10.1002/adfm.201904784}
  {\bibfield  {journal} {\bibinfo  {journal} {Advanced Functional Materials}\
  }\textbf {\bibinfo {volume} {n/a}},\ \bibinfo {pages} {1904784}}\BibitemShut
  {NoStop}%
\bibitem [{\citenamefont {Saffman}\ \emph {et~al.}(2010)\citenamefont
  {Saffman}, \citenamefont {Walker},\ and\ \citenamefont
  {M\o{}lmer}}]{Saffman_2010}%
  \BibitemOpen
  \bibfield  {author} {\bibinfo {author} {\bibfnamefont {M.}~\bibnamefont
  {Saffman}}, \bibinfo {author} {\bibfnamefont {T.~G.}\ \bibnamefont {Walker}},
  \ and\ \bibinfo {author} {\bibfnamefont {K.}~\bibnamefont {M\o{}lmer}},\
  }\bibfield  {title} {\enquote {\bibinfo {title} {{Quantum information with
  Rydberg atoms}},}\ }\href {\doibase 10.1103/RevModPhys.82.2313} {\bibfield
  {journal} {\bibinfo  {journal} {Rev. Mod. Phys.}\ }\textbf {\bibinfo {volume}
  {82}},\ \bibinfo {pages} {2313} (\bibinfo {year} {2010})}\BibitemShut
  {NoStop}%
\bibitem [{\citenamefont {Weimer}\ \emph {et~al.}(2010)\citenamefont {Weimer},
  \citenamefont {M{\"u}ller}, \citenamefont {Lesanovsky}, \citenamefont
  {Zoller},\ and\ \citenamefont {B{\"u}chler}}]{Weimer_2010}%
  \BibitemOpen
  \bibfield  {author} {\bibinfo {author} {\bibfnamefont {H.}~\bibnamefont
  {Weimer}}, \bibinfo {author} {\bibfnamefont {M.}~\bibnamefont {M{\"u}ller}},
  \bibinfo {author} {\bibfnamefont {I.}~\bibnamefont {Lesanovsky}}, \bibinfo
  {author} {\bibfnamefont {P.}~\bibnamefont {Zoller}}, \ and\ \bibinfo {author}
  {\bibfnamefont {H.~P.}\ \bibnamefont {B{\"u}chler}},\ }\bibfield  {title}
  {\enquote {\bibinfo {title} {{A Rydberg quantum simulator}},}\ }\href
  {\doibase 10.1038/nphys1614} {\bibfield  {journal} {\bibinfo  {journal}
  {Nature Physics}\ }\textbf {\bibinfo {volume} {6}},\ \bibinfo {pages} {382}
  (\bibinfo {year} {2010})}\BibitemShut {NoStop}%
\bibitem [{\citenamefont {Miroshnychenko}\ \emph {et~al.}(2006)\citenamefont
  {Miroshnychenko}, \citenamefont {Alt}, \citenamefont {Dotsenko},
  \citenamefont {F\"orster}, \citenamefont {Khudaverdyan}, \citenamefont
  {Rauschenbeutel},\ and\ \citenamefont {Meschede}}]{Miroshnychenko_2006}%
  \BibitemOpen
  \bibfield  {author} {\bibinfo {author} {\bibfnamefont {Y.}~\bibnamefont
  {Miroshnychenko}}, \bibinfo {author} {\bibfnamefont {W.}~\bibnamefont {Alt}},
  \bibinfo {author} {\bibfnamefont {I.}~\bibnamefont {Dotsenko}}, \bibinfo
  {author} {\bibfnamefont {L.}~\bibnamefont {F\"orster}}, \bibinfo {author}
  {\bibfnamefont {M.}~\bibnamefont {Khudaverdyan}}, \bibinfo {author}
  {\bibfnamefont {A.}~\bibnamefont {Rauschenbeutel}}, \ and\ \bibinfo {author}
  {\bibfnamefont {D.}~\bibnamefont {Meschede}},\ }\bibfield  {title} {\enquote
  {\bibinfo {title} {Precision preparation of strings of trapped neutral
  atoms},}\ }\href {\doibase 10.1088/1367-2630/8/9/191} {\bibfield  {journal}
  {\bibinfo  {journal} {New Journal of Physics}\ }\textbf {\bibinfo {volume}
  {8}},\ \bibinfo {pages} {191} (\bibinfo {year} {2006})}\BibitemShut {NoStop}%
\bibitem [{\citenamefont {Barredo}\ \emph {et~al.}(2016)\citenamefont
  {Barredo}, \citenamefont {de~L{\'e}s{\'e}leuc}, \citenamefont {Lienhard},
  \citenamefont {Lahaye},\ and\ \citenamefont {Browaeys}}]{Barredo_2016}%
  \BibitemOpen
  \bibfield  {author} {\bibinfo {author} {\bibfnamefont {D.}~\bibnamefont
  {Barredo}}, \bibinfo {author} {\bibfnamefont {S.}~\bibnamefont
  {de~L{\'e}s{\'e}leuc}}, \bibinfo {author} {\bibfnamefont {V.}~\bibnamefont
  {Lienhard}}, \bibinfo {author} {\bibfnamefont {T.}~\bibnamefont {Lahaye}}, \
  and\ \bibinfo {author} {\bibfnamefont {A.}~\bibnamefont {Browaeys}},\
  }\bibfield  {title} {\enquote {\bibinfo {title} {An atom-by-atom assembler of
  defect-free arbitrary two-dimensional atomic arrays},}\ }\href {\doibase
  10.1126/science.aah3778} {\bibfield  {journal} {\bibinfo  {journal}
  {Science}\ }\textbf {\bibinfo {volume} {354}},\ \bibinfo {pages} {1021}
  (\bibinfo {year} {2016})}\BibitemShut {NoStop}%
\bibitem [{\citenamefont {Endres}\ \emph {et~al.}(2016)\citenamefont {Endres},
  \citenamefont {Bernien}, \citenamefont {Keesling}, \citenamefont {Levine},
  \citenamefont {Anschuetz}, \citenamefont {Krajenbrink}, \citenamefont
  {Senko}, \citenamefont {Vuletic}, \citenamefont {Greiner},\ and\
  \citenamefont {Lukin}}]{Endres_2016}%
  \BibitemOpen
  \bibfield  {author} {\bibinfo {author} {\bibfnamefont {M.}~\bibnamefont
  {Endres}}, \bibinfo {author} {\bibfnamefont {H.}~\bibnamefont {Bernien}},
  \bibinfo {author} {\bibfnamefont {A.}~\bibnamefont {Keesling}}, \bibinfo
  {author} {\bibfnamefont {H.}~\bibnamefont {Levine}}, \bibinfo {author}
  {\bibfnamefont {E.~R.}\ \bibnamefont {Anschuetz}}, \bibinfo {author}
  {\bibfnamefont {A.}~\bibnamefont {Krajenbrink}}, \bibinfo {author}
  {\bibfnamefont {C.}~\bibnamefont {Senko}}, \bibinfo {author} {\bibfnamefont
  {V.}~\bibnamefont {Vuletic}}, \bibinfo {author} {\bibfnamefont
  {M.}~\bibnamefont {Greiner}}, \ and\ \bibinfo {author} {\bibfnamefont
  {M.~D.}\ \bibnamefont {Lukin}},\ }\bibfield  {title} {\enquote {\bibinfo
  {title} {Atom-by-atom assembly of defect-free one-dimensional cold atom
  arrays},}\ }\href {\doibase 10.1126/science.aah3752} {\bibfield  {journal}
  {\bibinfo  {journal} {Science}\ }\textbf {\bibinfo {volume} {354}},\ \bibinfo
  {pages} {1024} (\bibinfo {year} {2016})}\BibitemShut {NoStop}%
\bibitem [{\citenamefont {Kim}\ \emph {et~al.}(2016)\citenamefont {Kim},
  \citenamefont {Lee}, \citenamefont {Lee}, \citenamefont {Jo}, \citenamefont
  {Song},\ and\ \citenamefont {Ahn}}]{Kim_2016}%
  \BibitemOpen
  \bibfield  {author} {\bibinfo {author} {\bibfnamefont {H.}~\bibnamefont
  {Kim}}, \bibinfo {author} {\bibfnamefont {W.}~\bibnamefont {Lee}}, \bibinfo
  {author} {\bibfnamefont {H.}~\bibnamefont {Lee}}, \bibinfo {author}
  {\bibfnamefont {H.}~\bibnamefont {Jo}}, \bibinfo {author} {\bibfnamefont
  {Y.}~\bibnamefont {Song}}, \ and\ \bibinfo {author} {\bibfnamefont
  {J.}~\bibnamefont {Ahn}},\ }\bibfield  {title} {\enquote {\bibinfo {title}
  {In situ single-atom array synthesis using dynamic holographic optical
  tweezers},}\ }\href {\doibase 10.1038/ncomms13317} {\bibfield  {journal}
  {\bibinfo  {journal} {Nat. Commun.}\ }\textbf {\bibinfo {volume} {7}},\
  \bibinfo {pages} {13317} (\bibinfo {year} {2016})}\BibitemShut {NoStop}%
\bibitem [{\citenamefont {Barredo}\ \emph {et~al.}(2018)\citenamefont
  {Barredo}, \citenamefont {Lienhard}, \citenamefont {de~L{\'e}s{\'e}leuc},
  \citenamefont {Lahaye},\ and\ \citenamefont {Browaeys}}]{Barredo_2018}%
  \BibitemOpen
  \bibfield  {author} {\bibinfo {author} {\bibfnamefont {D.}~\bibnamefont
  {Barredo}}, \bibinfo {author} {\bibfnamefont {V.}~\bibnamefont {Lienhard}},
  \bibinfo {author} {\bibfnamefont {S.}~\bibnamefont {de~L{\'e}s{\'e}leuc}},
  \bibinfo {author} {\bibfnamefont {T.}~\bibnamefont {Lahaye}}, \ and\ \bibinfo
  {author} {\bibfnamefont {A.}~\bibnamefont {Browaeys}},\ }\bibfield  {title}
  {\enquote {\bibinfo {title} {Synthetic three-dimensional atomic structures
  assembled atom by atom},}\ }\href {\doibase 10.1038/s41586-018-0450-2}
  {\bibfield  {journal} {\bibinfo  {journal} {Nature}\ }\textbf {\bibinfo
  {volume} {561}},\ \bibinfo {pages} {79} (\bibinfo {year} {2018})}\BibitemShut
  {NoStop}%
\bibitem [{\citenamefont {Ohl~de Mello}\ \emph {et~al.}(2019)\citenamefont
  {Ohl~de Mello}, \citenamefont {Sch\"affner}, \citenamefont {Werkmann},
  \citenamefont {Preuschoff}, \citenamefont {Kohfahl}, \citenamefont
  {Schlosser},\ and\ \citenamefont {Birkl}}]{Mello2019}%
  \BibitemOpen
  \bibfield  {author} {\bibinfo {author} {\bibfnamefont {D.}~\bibnamefont
  {Ohl~de Mello}}, \bibinfo {author} {\bibfnamefont {D.}~\bibnamefont
  {Sch\"affner}}, \bibinfo {author} {\bibfnamefont {J.}~\bibnamefont
  {Werkmann}}, \bibinfo {author} {\bibfnamefont {T.}~\bibnamefont
  {Preuschoff}}, \bibinfo {author} {\bibfnamefont {L.}~\bibnamefont {Kohfahl}},
  \bibinfo {author} {\bibfnamefont {M.}~\bibnamefont {Schlosser}}, \ and\
  \bibinfo {author} {\bibfnamefont {G.}~\bibnamefont {Birkl}},\ }\bibfield
  {title} {\enquote {\bibinfo {title} {Defect-free assembly of {2D} clusters of
  more than 100 single-atom quantum systems},}\ }\href {\doibase
  10.1103/PhysRevLett.122.203601} {\bibfield  {journal} {\bibinfo  {journal}
  {Phys. Rev. Lett.}\ }\textbf {\bibinfo {volume} {122}},\ \bibinfo {pages}
  {203601} (\bibinfo {year} {2019})}\BibitemShut {NoStop}%
\bibitem [{\citenamefont {Browaeys}\ \emph {et~al.}(2016)\citenamefont
  {Browaeys}, \citenamefont {Barredo},\ and\ \citenamefont
  {Lahaye}}]{Browaeys_2016}%
  \BibitemOpen
  \bibfield  {author} {\bibinfo {author} {\bibfnamefont {A.}~\bibnamefont
  {Browaeys}}, \bibinfo {author} {\bibfnamefont {D.}~\bibnamefont {Barredo}}, \
  and\ \bibinfo {author} {\bibfnamefont {T.}~\bibnamefont {Lahaye}},\
  }\bibfield  {title} {\enquote {\bibinfo {title} {{Experimental investigations
  of dipole{\textendash}dipole interactions between a few Rydberg atoms}},}\
  }\href {\doibase 10.1088/0953-4075/49/15/152001} {\bibfield  {journal}
  {\bibinfo  {journal} {J. Phys. B}\ }\textbf {\bibinfo {volume} {49}},\
  \bibinfo {pages} {152001} (\bibinfo {year} {2016})}\BibitemShut {NoStop}%
\bibitem [{\citenamefont {Labuhn}\ \emph {et~al.}(2016)\citenamefont {Labuhn},
  \citenamefont {Barredo}, \citenamefont {Ravets}, \citenamefont
  {de~L{\'e}s{\'e}leuc}, \citenamefont {Macr{\`i}}, \citenamefont {Lahaye},\
  and\ \citenamefont {Browaeys}}]{Labuhn_2016}%
  \BibitemOpen
  \bibfield  {author} {\bibinfo {author} {\bibfnamefont {H.}~\bibnamefont
  {Labuhn}}, \bibinfo {author} {\bibfnamefont {D.}~\bibnamefont {Barredo}},
  \bibinfo {author} {\bibfnamefont {S.}~\bibnamefont {Ravets}}, \bibinfo
  {author} {\bibfnamefont {S.}~\bibnamefont {de~L{\'e}s{\'e}leuc}}, \bibinfo
  {author} {\bibnamefont {Macr{\`i}}}, \bibinfo {author} {\bibfnamefont
  {T.}~\bibnamefont {Lahaye}}, \ and\ \bibinfo {author} {\bibfnamefont
  {A.}~\bibnamefont {Browaeys}},\ }\bibfield  {title} {\enquote {\bibinfo
  {title} {{Tunable two-dimensional arrays of single Rydberg atoms for
  realizing quantum Ising models}},}\ }\href {\doibase 10.1038/nature18274}
  {\bibfield  {journal} {\bibinfo  {journal} {Nature}\ }\textbf {\bibinfo
  {volume} {534}},\ \bibinfo {pages} {667} (\bibinfo {year}
  {2016})}\BibitemShut {NoStop}%
\bibitem [{\citenamefont {Bernien}\ \emph {et~al.}(2017)\citenamefont
  {Bernien}, \citenamefont {Schwartz}, \citenamefont {Keesling}, \citenamefont
  {Levine}, \citenamefont {Omran}, \citenamefont {Pichler}, \citenamefont
  {Choi}, \citenamefont {Zibrov}, \citenamefont {Endres}, \citenamefont
  {Greiner}, \citenamefont {Vuletic},\ and\ \citenamefont
  {Lukin}}]{Bernien_2017}%
  \BibitemOpen
  \bibfield  {author} {\bibinfo {author} {\bibfnamefont {H.}~\bibnamefont
  {Bernien}}, \bibinfo {author} {\bibfnamefont {S.}~\bibnamefont {Schwartz}},
  \bibinfo {author} {\bibfnamefont {A.}~\bibnamefont {Keesling}}, \bibinfo
  {author} {\bibfnamefont {H.}~\bibnamefont {Levine}}, \bibinfo {author}
  {\bibfnamefont {A.}~\bibnamefont {Omran}}, \bibinfo {author} {\bibfnamefont
  {H.}~\bibnamefont {Pichler}}, \bibinfo {author} {\bibfnamefont
  {S.}~\bibnamefont {Choi}}, \bibinfo {author} {\bibfnamefont {A.~S.}\
  \bibnamefont {Zibrov}}, \bibinfo {author} {\bibfnamefont {M.}~\bibnamefont
  {Endres}}, \bibinfo {author} {\bibfnamefont {M.}~\bibnamefont {Greiner}},
  \bibinfo {author} {\bibfnamefont {V.}~\bibnamefont {Vuletic}}, \ and\
  \bibinfo {author} {\bibfnamefont {M.~D.}\ \bibnamefont {Lukin}},\ }\bibfield
  {title} {\enquote {\bibinfo {title} {Probing many-body dynamics on a 51-atom
  quantum simulator},}\ }\href {\doibase 10.1038/nature24622} {\bibfield
  {journal} {\bibinfo  {journal} {Nature}\ }\textbf {\bibinfo {volume} {551}},\
  \bibinfo {pages} {579} (\bibinfo {year} {2017})}\BibitemShut {NoStop}%
\bibitem [{\citenamefont {Kim}\ \emph {et~al.}(2018)\citenamefont {Kim},
  \citenamefont {Park}, \citenamefont {Kim}, \citenamefont {Sim},\ and\
  \citenamefont {Ahn}}]{Kim_2018}%
  \BibitemOpen
  \bibfield  {author} {\bibinfo {author} {\bibfnamefont {H.}~\bibnamefont
  {Kim}}, \bibinfo {author} {\bibfnamefont {Y.}~\bibnamefont {Park}}, \bibinfo
  {author} {\bibfnamefont {K.}~\bibnamefont {Kim}}, \bibinfo {author}
  {\bibfnamefont {H.-S.}\ \bibnamefont {Sim}}, \ and\ \bibinfo {author}
  {\bibfnamefont {J.}~\bibnamefont {Ahn}},\ }\bibfield  {title} {\enquote
  {\bibinfo {title} {Detailed balance of thermalization dynamics in
  {R}ydberg-atom quantum simulators},}\ }\href {\doibase
  10.1103/PhysRevLett.120.180502} {\bibfield  {journal} {\bibinfo  {journal}
  {Phys. Rev. Lett.}\ }\textbf {\bibinfo {volume} {120}},\ \bibinfo {pages}
  {180502} (\bibinfo {year} {2018})}\BibitemShut {NoStop}%
\bibitem [{\citenamefont {Barredo}\ \emph {et~al.}(2015)\citenamefont
  {Barredo}, \citenamefont {Labuhn}, \citenamefont {Ravets}, \citenamefont
  {Lahaye}, \citenamefont {Browaeys},\ and\ \citenamefont
  {Adams}}]{Barredo_2015}%
  \BibitemOpen
  \bibfield  {author} {\bibinfo {author} {\bibfnamefont {D.}~\bibnamefont
  {Barredo}}, \bibinfo {author} {\bibfnamefont {H.}~\bibnamefont {Labuhn}},
  \bibinfo {author} {\bibfnamefont {S.}~\bibnamefont {Ravets}}, \bibinfo
  {author} {\bibfnamefont {T.}~\bibnamefont {Lahaye}}, \bibinfo {author}
  {\bibfnamefont {A.}~\bibnamefont {Browaeys}}, \ and\ \bibinfo {author}
  {\bibfnamefont {C.~S.}\ \bibnamefont {Adams}},\ }\bibfield  {title} {\enquote
  {\bibinfo {title} {Coherent excitation transfer in a spin chain of three
  {R}ydberg atoms},}\ }\href {\doibase 10.1103/PhysRevLett.114.113002}
  {\bibfield  {journal} {\bibinfo  {journal} {Phys. Rev. Lett.}\ }\textbf
  {\bibinfo {volume} {114}},\ \bibinfo {pages} {113002} (\bibinfo {year}
  {2015})}\BibitemShut {NoStop}%
\bibitem [{\citenamefont {de~L{\'e}s{\'e}leuc}\ \emph
  {et~al.}(2019)\citenamefont {de~L{\'e}s{\'e}leuc}, \citenamefont {Lienhard},
  \citenamefont {Scholl}, \citenamefont {Barredo}, \citenamefont {Weber},
  \citenamefont {Lang}, \citenamefont {B{\"u}chler}, \citenamefont {Lahaye},\
  and\ \citenamefont {Browaeys}}]{deLeseleuc_2019}%
  \BibitemOpen
  \bibfield  {author} {\bibinfo {author} {\bibfnamefont {S.}~\bibnamefont
  {de~L{\'e}s{\'e}leuc}}, \bibinfo {author} {\bibfnamefont {V.}~\bibnamefont
  {Lienhard}}, \bibinfo {author} {\bibfnamefont {P.}~\bibnamefont {Scholl}},
  \bibinfo {author} {\bibfnamefont {D.}~\bibnamefont {Barredo}}, \bibinfo
  {author} {\bibfnamefont {S.}~\bibnamefont {Weber}}, \bibinfo {author}
  {\bibfnamefont {N.}~\bibnamefont {Lang}}, \bibinfo {author} {\bibfnamefont
  {H.~P.}\ \bibnamefont {B{\"u}chler}}, \bibinfo {author} {\bibfnamefont
  {T.}~\bibnamefont {Lahaye}}, \ and\ \bibinfo {author} {\bibfnamefont
  {A.}~\bibnamefont {Browaeys}},\ }\bibfield  {title} {\enquote {\bibinfo
  {title} {{Observation of a symmetry-protected topological phase of
  interacting bosons with Rydberg atoms}},}\ }\href {\doibase
  10.1126/science.aav9105} {\bibfield  {journal} {\bibinfo  {journal}
  {Science}\ }\textbf {\bibinfo {volume} {365}},\ \bibinfo {pages} {775}
  (\bibinfo {year} {2019})}\BibitemShut {NoStop}%
\bibitem [{\citenamefont {Syzranov}\ \emph {et~al.}(2014)\citenamefont
  {Syzranov}, \citenamefont {Wall}, \citenamefont {Gurarie},\ and\
  \citenamefont {Rey}}]{Syzranov2014}%
  \BibitemOpen
  \bibfield  {author} {\bibinfo {author} {\bibfnamefont {S.~V.}\ \bibnamefont
  {Syzranov}}, \bibinfo {author} {\bibfnamefont {M.~L.}\ \bibnamefont {Wall}},
  \bibinfo {author} {\bibfnamefont {V.}~\bibnamefont {Gurarie}}, \ and\
  \bibinfo {author} {\bibfnamefont {A.~M.}\ \bibnamefont {Rey}},\ }\bibfield
  {title} {\enquote {\bibinfo {title} {Spin--orbital dynamics in a system of
  polar molecules},}\ }\href {\doibase 10.1038/ncomms6391} {\bibfield
  {journal} {\bibinfo  {journal} {Nature Communications}\ }\textbf {\bibinfo
  {volume} {5}},\ \bibinfo {pages} {5391} (\bibinfo {year} {2014})}\BibitemShut
  {NoStop}%
\bibitem [{\citenamefont {Peter}\ \emph {et~al.}(2015)\citenamefont {Peter},
  \citenamefont {Yao}, \citenamefont {Lang}, \citenamefont {Huber},
  \citenamefont {Lukin},\ and\ \citenamefont {B\"uchler}}]{Peter_2015}%
  \BibitemOpen
  \bibfield  {author} {\bibinfo {author} {\bibfnamefont {D.}~\bibnamefont
  {Peter}}, \bibinfo {author} {\bibfnamefont {N.~Y.}\ \bibnamefont {Yao}},
  \bibinfo {author} {\bibfnamefont {N.}~\bibnamefont {Lang}}, \bibinfo {author}
  {\bibfnamefont {S.~D.}\ \bibnamefont {Huber}}, \bibinfo {author}
  {\bibfnamefont {M.~D.}\ \bibnamefont {Lukin}}, \ and\ \bibinfo {author}
  {\bibfnamefont {H.~P.}\ \bibnamefont {B\"uchler}},\ }\bibfield  {title}
  {\enquote {\bibinfo {title} {{Topological bands with a Chern number $C=2$ by
  dipolar exchange interactions}},}\ }\href {\doibase
  10.1103/PhysRevA.91.053617} {\bibfield  {journal} {\bibinfo  {journal} {Phys.
  Rev. A}\ }\textbf {\bibinfo {volume} {91}},\ \bibinfo {pages} {053617}
  (\bibinfo {year} {2015})}\BibitemShut {NoStop}%
\bibitem [{\citenamefont {Kiffner}\ \emph {et~al.}(2017)\citenamefont
  {Kiffner}, \citenamefont {O'Brien},\ and\ \citenamefont
  {Jaksch}}]{Kiffner_2017}%
  \BibitemOpen
  \bibfield  {author} {\bibinfo {author} {\bibfnamefont {M.}~\bibnamefont
  {Kiffner}}, \bibinfo {author} {\bibfnamefont {E.}~\bibnamefont {O'Brien}}, \
  and\ \bibinfo {author} {\bibfnamefont {D.}~\bibnamefont {Jaksch}},\
  }\bibfield  {title} {\enquote {\bibinfo {title} {Topological spin models in
  {R}ydberg lattices},}\ }\href {\doibase 10.1007/s00340-016-6596-4} {\bibfield
   {journal} {\bibinfo  {journal} {Applied Physics B}\ }\textbf {\bibinfo
  {volume} {123}},\ \bibinfo {pages} {46} (\bibinfo {year} {2017})}\BibitemShut
  {NoStop}%
\bibitem [{\citenamefont {Weber}\ \emph {et~al.}(2018)\citenamefont {Weber},
  \citenamefont {de~L{\'{e}}s{\'{e}}leuc}, \citenamefont {Lienhard},
  \citenamefont {Barredo}, \citenamefont {Lahaye}, \citenamefont {Browaeys},\
  and\ \citenamefont {B{\"u}chler}}]{Weber_2018}%
  \BibitemOpen
  \bibfield  {author} {\bibinfo {author} {\bibfnamefont {S.}~\bibnamefont
  {Weber}}, \bibinfo {author} {\bibfnamefont {S.}~\bibnamefont
  {de~L{\'{e}}s{\'{e}}leuc}}, \bibinfo {author} {\bibfnamefont
  {V.}~\bibnamefont {Lienhard}}, \bibinfo {author} {\bibfnamefont
  {D.}~\bibnamefont {Barredo}}, \bibinfo {author} {\bibfnamefont
  {T.}~\bibnamefont {Lahaye}}, \bibinfo {author} {\bibfnamefont
  {A.}~\bibnamefont {Browaeys}}, \ and\ \bibinfo {author} {\bibfnamefont
  {H.~P.}\ \bibnamefont {B{\"u}chler}},\ }\bibfield  {title} {\enquote
  {\bibinfo {title} {{Topologically protected edge states in small Rydberg
  systems}},}\ }\href {\doibase 10.1088/2058-9565/aaca47} {\bibfield  {journal}
  {\bibinfo  {journal} {Quantum Science and Technology}\ }\textbf {\bibinfo
  {volume} {3}},\ \bibinfo {pages} {044001} (\bibinfo {year}
  {2018})}\BibitemShut {NoStop}%
\bibitem [{\citenamefont {Wiese}(2013)}]{Wiese2013}%
  \BibitemOpen
  \bibfield  {author} {\bibinfo {author} {\bibfnamefont {U.-J.}\ \bibnamefont
  {Wiese}},\ }\bibfield  {title} {\enquote {\bibinfo {title} {Ultracold quantum
  gases and lattice systems: quantum simulation of lattice gauge theories},}\
  }\href {\doibase 10.1002/andp.201300104} {\bibfield  {journal} {\bibinfo
  {journal} {Annalen der Physik}\ }\textbf {\bibinfo {volume} {525}},\ \bibinfo
  {pages} {777} (\bibinfo {year} {2013})}\BibitemShut {NoStop}%
\bibitem [{\citenamefont {G{\"o}rg}\ \emph {et~al.}(2019)\citenamefont
  {G{\"o}rg}, \citenamefont {Sandholzer}, \citenamefont {Minguzzi},
  \citenamefont {Desbuquois}, \citenamefont {Messer},\ and\ \citenamefont
  {Esslinger}}]{Esslinger_2019}%
  \BibitemOpen
  \bibfield  {author} {\bibinfo {author} {\bibfnamefont {F.}~\bibnamefont
  {G{\"o}rg}}, \bibinfo {author} {\bibfnamefont {K.}~\bibnamefont
  {Sandholzer}}, \bibinfo {author} {\bibfnamefont {J.}~\bibnamefont
  {Minguzzi}}, \bibinfo {author} {\bibfnamefont {R.}~\bibnamefont
  {Desbuquois}}, \bibinfo {author} {\bibfnamefont {M.}~\bibnamefont {Messer}},
  \ and\ \bibinfo {author} {\bibfnamefont {T.}~\bibnamefont {Esslinger}},\
  }\bibfield  {title} {\enquote {\bibinfo {title} {{Realization of
  density-dependent Peierls phases to engineer quantized gauge fields coupled
  to ultracold matter}},}\ }\href {\doibase 10.1038/s41567-019-0615-4}
  {\bibfield  {journal} {\bibinfo  {journal} {Nature Physics}\ }\textbf
  {\bibinfo {volume} {15}},\ \bibinfo {pages} {1161} (\bibinfo {year}
  {2019})}\BibitemShut {NoStop}%
\bibitem [{\citenamefont {Clark}\ \emph {et~al.}(2018)\citenamefont {Clark},
  \citenamefont {Anderson}, \citenamefont {Feng}, \citenamefont {Gaj},
  \citenamefont {Levin},\ and\ \citenamefont {Chin}}]{Clark2018}%
  \BibitemOpen
  \bibfield  {author} {\bibinfo {author} {\bibfnamefont {L.~W.}\ \bibnamefont
  {Clark}}, \bibinfo {author} {\bibfnamefont {B.~M.}\ \bibnamefont {Anderson}},
  \bibinfo {author} {\bibfnamefont {L.}~\bibnamefont {Feng}}, \bibinfo {author}
  {\bibfnamefont {A.}~\bibnamefont {Gaj}}, \bibinfo {author} {\bibfnamefont
  {K.}~\bibnamefont {Levin}}, \ and\ \bibinfo {author} {\bibfnamefont
  {C.}~\bibnamefont {Chin}},\ }\bibfield  {title} {\enquote {\bibinfo {title}
  {Observation of density-dependent gauge fields in a {B}ose-{E}instein
  condensate based on micromotion control in a shaken two-dimensional
  lattice},}\ }\href {\doibase 10.1103/PhysRevLett.121.030402} {\bibfield
  {journal} {\bibinfo  {journal} {Phys. Rev. Lett.}\ }\textbf {\bibinfo
  {volume} {121}},\ \bibinfo {pages} {030402} (\bibinfo {year}
  {2018})}\BibitemShut {NoStop}%
\bibitem [{\citenamefont {Fradkin}(1991)}]{Fradkin91}%
  \BibitemOpen
  \bibfield  {author} {\bibinfo {author} {\bibfnamefont {E.}~\bibnamefont
  {Fradkin}},\ }\href@noop {} {\emph {\bibinfo {title} {Field Theories of
  Condensed Matter Systems}}}\ (\bibinfo  {publisher} {Addison-Wesley, Redwood
  City, CA},\ \bibinfo {year} {1991})\BibitemShut {NoStop}%
\bibitem [{\citenamefont {Weber}\ \emph {et~al.}(2017)\citenamefont {Weber},
  \citenamefont {Tresp}, \citenamefont {Menke}, \citenamefont {Urvoy},
  \citenamefont {Firstenberg}, \citenamefont {B\"uchler},\ and\ \citenamefont
  {Hofferberth}}]{Weber_2017}%
  \BibitemOpen
  \bibfield  {author} {\bibinfo {author} {\bibfnamefont {S.}~\bibnamefont
  {Weber}}, \bibinfo {author} {\bibfnamefont {C.}~\bibnamefont {Tresp}},
  \bibinfo {author} {\bibfnamefont {H.}~\bibnamefont {Menke}}, \bibinfo
  {author} {\bibfnamefont {A.}~\bibnamefont {Urvoy}}, \bibinfo {author}
  {\bibfnamefont {O.}~\bibnamefont {Firstenberg}}, \bibinfo {author}
  {\bibfnamefont {H.~P.}\ \bibnamefont {B\"uchler}}, \ and\ \bibinfo {author}
  {\bibfnamefont {S.}~\bibnamefont {Hofferberth}},\ }\bibfield  {title}
  {\enquote {\bibinfo {title} {{Calculation of Rydberg interaction
  potentials}},}\ }\href {\doibase 10.1088/1361-6455/aa743a} {\bibfield
  {journal} {\bibinfo  {journal} {J. Phys. B}\ }\textbf {\bibinfo {volume}
  {50}},\ \bibinfo {pages} {133001} (\bibinfo {year} {2017})}\BibitemShut
  {NoStop}%
\bibitem [{\citenamefont {de~L\'es\'eleuc}\ \emph {et~al.}(2017)\citenamefont
  {de~L\'es\'eleuc}, \citenamefont {Barredo}, \citenamefont {Lienhard},
  \citenamefont {Browaeys},\ and\ \citenamefont {Lahaye}}]{deLeseleuc_2017}%
  \BibitemOpen
  \bibfield  {author} {\bibinfo {author} {\bibfnamefont {S.}~\bibnamefont
  {de~L\'es\'eleuc}}, \bibinfo {author} {\bibfnamefont {D.}~\bibnamefont
  {Barredo}}, \bibinfo {author} {\bibfnamefont {V.}~\bibnamefont {Lienhard}},
  \bibinfo {author} {\bibfnamefont {A.}~\bibnamefont {Browaeys}}, \ and\
  \bibinfo {author} {\bibfnamefont {T.}~\bibnamefont {Lahaye}},\ }\bibfield
  {title} {\enquote {\bibinfo {title} {Optical control of the resonant
  dipole-dipole interaction between {R}ydberg atoms},}\ }\href {\doibase
  10.1103/PhysRevLett.119.053202} {\bibfield  {journal} {\bibinfo  {journal}
  {Phys. Rev. Lett.}\ }\textbf {\bibinfo {volume} {119}},\ \bibinfo {pages}
  {053202} (\bibinfo {year} {2017})}\BibitemShut {NoStop}%
\bibitem [{\citenamefont {Zhu}\ and\ \citenamefont {Wang}(1996)}]{Zhu96}%
  \BibitemOpen
  \bibfield  {author} {\bibinfo {author} {\bibfnamefont {J.-X.}\ \bibnamefont
  {Zhu}}\ and\ \bibinfo {author} {\bibfnamefont {Z.~D.}\ \bibnamefont {Wang}},\
  }\bibfield  {title} {\enquote {\bibinfo {title} {Topological effects
  associated with fractional statistics in one-dimensional mesoscopic rings},}\
  }\href {\doibase 10.1103/PhysRevA.53.600} {\bibfield  {journal} {\bibinfo
  {journal} {Phys. Rev. A}\ }\textbf {\bibinfo {volume} {53}},\ \bibinfo
  {pages} {600} (\bibinfo {year} {1996})}\BibitemShut {NoStop}%
\bibitem [{\citenamefont {Kundu}(1999)}]{kundu_1999}%
  \BibitemOpen
  \bibfield  {author} {\bibinfo {author} {\bibfnamefont {A.}~\bibnamefont
  {Kundu}},\ }\bibfield  {title} {\enquote {\bibinfo {title} {Exact solution of
  double $\ensuremath{\delta}$ function {B}ose gas through an interacting anyon
  gas},}\ }\href {\doibase 10.1103/PhysRevLett.83.1275} {\bibfield  {journal}
  {\bibinfo  {journal} {Phys. Rev. Lett.}\ }\textbf {\bibinfo {volume} {83}},\
  \bibinfo {pages} {1275} (\bibinfo {year} {1999})}\BibitemShut {NoStop}%
\bibitem [{\citenamefont {Keilmann}\ \emph {et~al.}(2011)\citenamefont
  {Keilmann}, \citenamefont {Lanzmich}, \citenamefont {McCulloch},\ and\
  \citenamefont {Roncaglia}}]{Keilmann2011}%
  \BibitemOpen
  \bibfield  {author} {\bibinfo {author} {\bibfnamefont {T.}~\bibnamefont
  {Keilmann}}, \bibinfo {author} {\bibfnamefont {S.}~\bibnamefont {Lanzmich}},
  \bibinfo {author} {\bibfnamefont {I.}~\bibnamefont {McCulloch}}, \ and\
  \bibinfo {author} {\bibfnamefont {M.}~\bibnamefont {Roncaglia}},\ }\bibfield
  {title} {\enquote {\bibinfo {title} {{Statistically induced phase transitions
  and anyons in 1D optical lattices}},}\ }\href {\doibase 10.1038/ncomms1353}
  {\bibfield  {journal} {\bibinfo  {journal} {Nature Communications}\ }\textbf
  {\bibinfo {volume} {2}},\ \bibinfo {pages} {361} (\bibinfo {year}
  {2011})}\BibitemShut {NoStop}%
\bibitem [{\citenamefont {Greschner}\ and\ \citenamefont
  {Santos}(2015)}]{Greschner2015}%
  \BibitemOpen
  \bibfield  {author} {\bibinfo {author} {\bibfnamefont {S.}~\bibnamefont
  {Greschner}}\ and\ \bibinfo {author} {\bibfnamefont {L.}~\bibnamefont
  {Santos}},\ }\bibfield  {title} {\enquote {\bibinfo {title} {{Anyon Hubbard
  Model in One-Dimensional Optical Lattices}},}\ }\href {\doibase
  10.1103/PhysRevLett.115.053002} {\bibfield  {journal} {\bibinfo  {journal}
  {Phys. Rev. Lett.}\ }\textbf {\bibinfo {volume} {115}},\ \bibinfo {pages}
  {053002} (\bibinfo {year} {2015})}\BibitemShut {NoStop}%
\bibitem [{\citenamefont {Haldane}(1988)}]{Haldane1988}%
  \BibitemOpen
  \bibfield  {author} {\bibinfo {author} {\bibfnamefont {F.~D.~M.}\
  \bibnamefont {Haldane}},\ }\bibfield  {title} {\enquote {\bibinfo {title}
  {{Model for a Quantum Hall Effect without Landau Levels: Condensed-Matter
  Realization of the Parity Anomaly}},}\ }\href {\doibase
  10.1103/PhysRevLett.61.2015} {\bibfield  {journal} {\bibinfo  {journal}
  {Phys. Rev. Lett.}\ }\textbf {\bibinfo {volume} {61}},\ \bibinfo {pages}
  {2015} (\bibinfo {year} {1988})}\BibitemShut {NoStop}%
\bibitem [{InP()}]{InPrep}%
  \BibitemOpen
  \href@noop {} {}\bibinfo {note} {In preparation}\BibitemShut {NoStop}%
\bibitem [{\citenamefont {Wang}\ \emph {et~al.}(2011)\citenamefont {Wang},
  \citenamefont {Gu}, \citenamefont {Gong},\ and\ \citenamefont
  {Sheng}}]{Wan11}%
  \BibitemOpen
  \bibfield  {author} {\bibinfo {author} {\bibfnamefont {Y.-F.}\ \bibnamefont
  {Wang}}, \bibinfo {author} {\bibfnamefont {Z.-C.}\ \bibnamefont {Gu}},
  \bibinfo {author} {\bibfnamefont {C.-D.}\ \bibnamefont {Gong}}, \ and\
  \bibinfo {author} {\bibfnamefont {D.~N.}\ \bibnamefont {Sheng}},\ }\bibfield
  {title} {\enquote {\bibinfo {title} {{Fractional Quantum Hall Effect of
  Hard-Core Bosons in Topological Flat Bands}},}\ }\href {\doibase
  10.1103/PhysRevLett.107.146803} {\bibfield  {journal} {\bibinfo  {journal}
  {Phys. Rev. Lett.}\ }\textbf {\bibinfo {volume} {107}},\ \bibinfo {pages}
  {146803} (\bibinfo {year} {2011})}\BibitemShut {NoStop}%
\bibitem [{\citenamefont {Wang}\ \emph {et~al.}(2012)\citenamefont {Wang},
  \citenamefont {Yao}, \citenamefont {Gong},\ and\ \citenamefont
  {Sheng}}]{Wan12}%
  \BibitemOpen
  \bibfield  {author} {\bibinfo {author} {\bibfnamefont {Y.-F.}\ \bibnamefont
  {Wang}}, \bibinfo {author} {\bibfnamefont {H.}~\bibnamefont {Yao}}, \bibinfo
  {author} {\bibfnamefont {C.-D.}\ \bibnamefont {Gong}}, \ and\ \bibinfo
  {author} {\bibfnamefont {D.~N.}\ \bibnamefont {Sheng}},\ }\bibfield  {title}
  {\enquote {\bibinfo {title} {{Fractional quantum Hall effect in topological
  flat bands with Chern number two}},}\ }\href {\doibase
  10.1103/PhysRevB.86.201101} {\bibfield  {journal} {\bibinfo  {journal} {Phys.
  Rev. B}\ }\textbf {\bibinfo {volume} {86}},\ \bibinfo {pages} {201101}
  (\bibinfo {year} {2012})}\BibitemShut {NoStop}%
\bibitem [{\citenamefont {de~L\'es\'eleuc}\ \emph {et~al.}(2018)\citenamefont
  {de~L\'es\'eleuc}, \citenamefont {Barredo}, \citenamefont {Lienhard},
  \citenamefont {Browaeys},\ and\ \citenamefont {Lahaye}}]{deLeseleuc_2018}%
  \BibitemOpen
  \bibfield  {author} {\bibinfo {author} {\bibfnamefont {S.}~\bibnamefont
  {de~L\'es\'eleuc}}, \bibinfo {author} {\bibfnamefont {D.}~\bibnamefont
  {Barredo}}, \bibinfo {author} {\bibfnamefont {V.}~\bibnamefont {Lienhard}},
  \bibinfo {author} {\bibfnamefont {A.}~\bibnamefont {Browaeys}}, \ and\
  \bibinfo {author} {\bibfnamefont {T.}~\bibnamefont {Lahaye}},\ }\bibfield
  {title} {\enquote {\bibinfo {title} {{Analysis of imperfections in the
  coherent optical excitation of single atoms to Rydberg states}},}\ }\href
  {\doibase 10.1103/PhysRevA.97.053803} {\bibfield  {journal} {\bibinfo
  {journal} {Phys. Rev. A}\ }\textbf {\bibinfo {volume} {97}},\ \bibinfo
  {pages} {053803} (\bibinfo {year} {2018})}\BibitemShut {NoStop}%
\bibitem [{\citenamefont {Zee}(1990)}]{zee_1990}%
  \BibitemOpen
  \bibfield  {author} {\bibinfo {author} {\bibfnamefont {A.}~\bibnamefont
  {Zee}},\ }\bibfield  {title} {\enquote {\bibinfo {title} {Semionics: {A}
  theory of superconductivity based on fractional quantum statistics},}\ }in\
  \href@noop {} {\emph {\bibinfo {booktitle} {25th International Conference on
  High-Energy Physics}}},\ Vol.\ \bibinfo {volume} {900802}\ (\bibinfo {year}
  {1990})\ pp.\ \bibinfo {pages} {206--212}\BibitemShut {NoStop}%
\end{thebibliography}

%

\end{document}